\documentstyle[12pt,epsfig]{article}  
\newlength{\dinwidth}                       
\newlength{\dinmargin}                      
\def\lsim{\mathrel{\rlap{\lower4pt\hbox{\hskip1pt$\sim$}}
    \raise1pt\hbox{$<$}}}                
\def\gsim{\mathrel{\rlap{\lower4pt\hbox{\hskip1pt$\sim$}}
    \raise1pt\hbox{$>$}}}                
\begin{document}
\vspace*{1cm}
\begin{center}  \begin{Large} \begin{bf}
Nuclear beams in HERA\\
  \end{bf}  \end{Large}
  \vspace*{5mm}
  \begin{large}
M.Arneodo$^a$, A.Bialas$^{b}$, M.W.Krasny$^{c}$, T.Sloan$^d$ 
 and M. Strikman$^e$\\ 
  \end{large}
\end{center}
$^a$ Universit\`a di Torino,  
     I-10125 and INFN Cosenza, Italy\\
$^b$ Institute of Physics, Jagellonian University, Cracow, 
      Poland\\
$^c$ LPNHE, Universit\'es Paris VI and VII, IN2P3-CNRS, Paris, France\\
$^d$ School of Physics and Chemistry, University of Lancaster, 
      Lancaster LA1 4YB, UK\\
$^e$ Pennsylvania State University, University Park, PA 16802, USA\\
\begin{quotation}
\noindent
{\bf Abstract:}
A study has been made of the physics interest and feasibility of experiments 
with nuclear beams in HERA.  It is shown that such experiments widen 
considerably the horizon for probing QCD compared to that from free nucleon 
targets.  In addition there is some sensitivity to physics beyond the 
standard model.  Hence the option to include circulating 
nuclear beams in HERA allows a wide range of physics processes to be studied 
and understood.  
\end{quotation}
(Submitted to the Proceedings of the Workshop on Future Physics at HERA)
\section{Introduction}
\label{mainintroduction}

The successes of QCD in describing {\it inclusive} perturbative phenomena 
have moved the focus of investigations to new frontiers.  Three fundamental 
questions to be resolved are the space-time structure of high-energy strong 
interactions, the QCD dynamics in the nonlinear, small coupling domain and 
the QCD dynamics of interactions of fast, compact colour singlet systems. 

The study of electron-nucleus scattering at HERA allows a new 
regime to be probed experimentally for the first time.  This 
is the regime in which the virtual photon interacts coherently with 
all the nucleons at a given impact parameter.
In the rest frame of the nucleus this can be visualized in terms of  the  
propagation of a small $q \bar q$ pair in high density gluon fields 
over much larger distances than is possible with free nucleons. 
In the Breit frame it corresponds to the fact that small $x$ partons 
cannot be localized longitudinally to better than the size of the  nucleus.  
Thus low $x$  partons from different nucleons 
overlap  spatially creating much larger parton densities than in the free 
nucleon case. This leads to 
{\bf a large amplification of the nonlinear effects expected in QCD 
at small $x$.}
\unboldmath
The HERA $ep$ data have confirmed the rapid increase of the parton 
densities in the small $x$ limit predicted by perturbative QCD. 
However the limited $x$ 
range available at HERA makes it difficult to distinguish between the 
predictions of the DGLAP evolution equations and the BFKL-type dynamics. 
Moreover, the nonlinear effects expected at  small $x$  are relatively 
small in $ep$ scattering in the HERA kinematic domain and it may  be 
necessary to reduce $x$ by at least one order of magnitude to observe 
unambiguously such  effects.  However, the amplification obtained with heavy 
\boldmath
nuclear targets {\bf allows an effective reduction of about two orders of 
magnitude in $x$} making it feasible to explore such nonlinear effects at 
\unboldmath
the energies available at HERA.  The question of nonlinear effects is 
one of the most fundamental in QCD.  It is crucial for understanding 
the kind of dynamics which would slow down and eventually stop the rapid 
growth of the cross section (or the structure function,$F_2$) at small $x$.  
It is also essential in order to understand down to what values of  
$x$ the decomposition of the cross section into terms with different  
powers of $1 \over Q^2$ remains effective. It is important for the  
understanding of the relationship between hard and soft 
physics.  One can also study the dynamics of QCD at high densities 
and at zero temperatures raising questions complementary to those 
addressed in the search for a quark-gluon plasma in high-energy  
heavy ion collisions.

Deep inelastic scattering from nuclei provides also a number of ways to 
probe
{\bf the dynamics of high-energy interactions of small colour singlet 
systems}.  This issue started from the work of Gribov \cite{Gribov69} 
who demonstrated the following paradox.  If one makes the natural (in soft 
physics) assumption that at high energies any hadron interacts with a 
heavy nucleus with cross section $2 \pi R_A^2$ (corresponding to interaction 
with a black body), Bjorken scaling at small $x$ is grossly violated -- 
$\sigma_{\gamma^*A} \propto \ln Q^2$ instead of $1 \over Q^2$.  
To preserve scaling, Bjorken suggested, using parton model arguments,  that 
only configurations with small $p_t \le p_{t0}$ are involved in the 
interaction (the Aligned Jet Model) \cite{BJ71}.  However, in perturbative 
QCD  Bjorken's assumption does not hold -- large $p_t$ configurations 
interact with finite though small cross sections (colour screening), 
which however increase rapidly  with incident energy due to the 
increase of the gluon density with decreasing  $x$.
 Hence again one is faced with a fundamental
question which can only be answered experimentally: 
{\it  Can} {\bf  small} {\it colour singlets interact with hadrons
 with cross sections comparable to that of normal hadrons?} At HERA
 one can both establish the $x,Q^2$ range where the cross section 
 of small colour singlets is small -- {\it colour transparency}, and 
look for the  onset of the new regime of large cross sections,  
 {\it perturbative colour opacity}.

Another fundamental question to be addressed is {\bf the propagation of quarks 
through nuclear matter.}  At large energies perturbative QCD leads to  
the analogue of the Landau-Migdal-Pomeranchuk effect in quantum 
electrodynamics.  In particular Baier et al.  \cite{Baier} find a 
highly nontrivial dependence of the energy loss on the distance, $L$, 
travelled by a parton in a nuclear medium:  the loss  instead of being 
$\propto L$  is  $\propto L^2$.  Several manifestations of this phenomenon 
can be studied at HERA.

There is also  an {\it important connection to heavy ion physics.}  Study of  
$eA$ scattering at  HERA  would be important for the analysis  of heavy 
ion collisions at the LHC and RHIC.  Measurements of gluon shadowing at 
small $x$ are necessary for a reliable interpretation of the high $p_t$ 
jet rates at the LHC.  In addition, the study of parton propagation in 
nuclear media is important for the analysis of jet quenching phenomena, 
which may be one of the most direct global signals of the formation 
of a quark-gluon plasma.   

\begin{figure}
\vspace{-1.1cm}
\begin{center}
\leavevmode
\epsfig{file=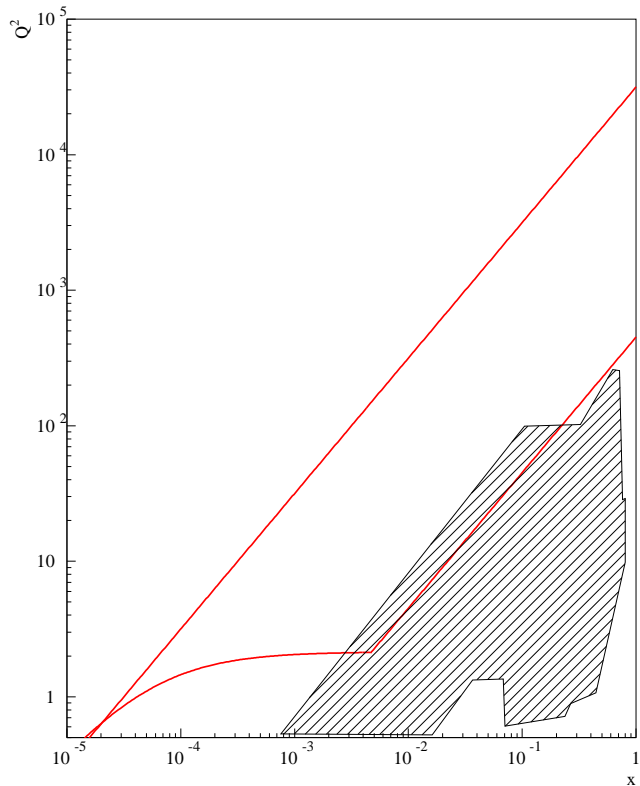,height=10cm,bbllx=19 ,bblly=142 , bburx=577 , bbury=700}
\end{center}
\caption{
The kinematic region covered by experiments at HERA compared to fixed target
 data (shaded region).  
}
\label{kinem}
\end{figure}

Current fixed target data on lepton-nucleus scattering only touch the surface 
of all these effects due to the limited $Q^2$ range of the data at small $x$.  
Indeed the $Q^2$ range of these data is too small to distinguish 
the contribution of the vector meson dominance behaviour of  
the photon from its hard QCD  behaviour at small $x$.  The range of $x$ and 
$Q^2$ in experiments with nuclei at HERA compared to the fixed target 
experiments is shown in Fig.~\ref{kinem}.  
It can be seen that at HERA the kinematic range will be extended well into 
the deep inelastic scattering region.  

To address the questions discussed above we identify
the primary experimental programme for nuclei in HERA as:  

$\bullet$
\boldmath 
{\bf Study 
of the $x$ and $Q^2$ dependence of nuclear shadowing over a wide $Q^2$
range.}
\unboldmath
  This will allow 
the processes limiting the growth of $F_2$ as $x$ tends to zero to be 
studied in detail.  

$\bullet$ {\bf To establish the difference between the gluon distributions of 
bound and free nucleons.}  This will allow the part played by gluon fusion in 
the shadowing process to be studied directly.  

$\bullet$ {\bf Study of diffractive processes:}  
to see if the pomeron generated by 
nuclei shows any difference from that generated by free nucleons.  Processes 
such as vector meson production can also be used to search for colour 
transparency.  

$\bullet$ {\bf Study of hadronic final states.}  This allows the 
propagation of partons in the nuclear medium to be studied as well as the 
multiplicity fluctuations discussed later.  

The proceedings of Working Group 8 are organised as follows.  First we give 
an experimental overview in which we demonstrate the feasibility of 
carrying out this experimental programme.  Then we give a theoretical 
overview in which we explain the relevance of the programme to QCD.
Finally, we give the detailed contributions on different topics which 
demonstrate the depth of the physics interest.  
The proposed  measurements in the main will be possible with the 
existing detectors H1 and ZEUS measuring down to low $Q^2$ and 
with luminosities at the level of 1-10 pb$^{-1}$ per nucleon.  
The contribution of Chwastowski and Krasny \cite{KRASNY} shows that if the 
detectors could extend their rapidity coverages various experiments of 
interest to nuclear structure physicists become feasible.

\section{Experimental Overview}
\subsection{Introduction}

In the following subsections the feasibility of the measurements 
defined above as the primary experimental programme is 
investigated.  The nuclear targets should each have $Z/A$ of 1/2. 
Hence the energy of each nucleon in a deep inelastic collision will 
 be half that of the HERA proton energy of 820 GeV i.e. 410 GeV.  
The electron energy is assumed to have the standard value of 27.6 GeV.  
 We show that most 
studies can be carried out with the existing detectors requiring 
luminosities between 1-10~pb$^{-1}$ per nucleon.  
While the possibility of storing heavy nuclei up to Sn and Pb is very 
attractive, a program covering the light isoscalar nuclei (D, $^4$He, 
C, S) would by itself have a major discovery potential.  The necessary 
radiative corrections are described in the contributions of 
Kurek \cite{Kurek} and of Akushevich and Spiesberger \cite{AANDS}
who show that such corrections 
can be kept under control using suitable cuts on the data.  
In most of the experiments it is proposed to measure ratios of yields.  Hence, 
in order to minimise systematic errors, it is desirable to store 
different nuclei in HERA simultaneously.  Otherwise frequent 
changes of the stored nucleus in the beam will be necessary.  

\subsection{Shadowing Measurements Using Nuclei in HERA}
\label{The Accuracy of Shadowing Measurements Using Nuclei in HERA} 

 The accuracy of shadowing measurements for an experiment in which nuclear
targets are stored in HERA is calculated.  It is shown   
that for luminosities 
of 2 pb$^{-1}$ per nucleon the measurements would extend considerably the 
accuracy and range of the existing data.  

\begin{figure}
\begin{center}
\leavevmode
\epsfig{file=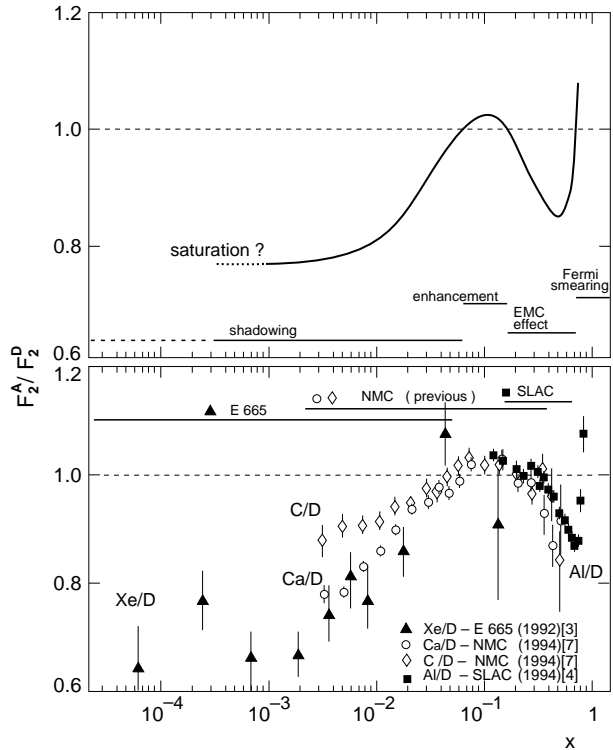,height=10cm}
\end{center}
\vspace{-0.5cm}
\caption{
Partial compilation of results for $F_2^{A}/F_2^D$
(from~\protect\cite{nmc_li}). Note that below $x \approx 0.01$, the 
average $Q^2$ value of the data is smaller than 2~GeV$^2$.
}
\label{NMCcompilation}
\end{figure}

The $x$ dependence of the differences in the nucleon structure function 
between bound and free nucleons has been well measured in recent
 years  in fixed target experiments~\cite{nmc_rean,nmc_li,e665s,
nmc_adep,nmc_q2dep} following the discovery of the differences in the 
1980s \cite{emc,emc1,emcna28,SLAC,BCDMS}.   
The present experimental situation on this $x$ dependence 
is briefly summarised in Fig.~\ref{NMCcompilation}.   However,  
the $Q^2$ dependence is not well measured.  In the shadowing 
region the data are at such low $Q^2$ values that  they  are arguably 
not even in the deep inelastic regime.  

There are many different models~\cite{reviews}
for the effects in the different 
regions shown in Fig.~\ref{NMCcompilation} which are all compatible with 
the existing data.  In the shadowing region the $Q^2$ range of the data 
is insufficient to separate the different contributions from the vector 
dominance behaviour of the photon and QCD effects such as parton fusion.    
Measurements over the extended $x$ and  $Q^2$  ranges,  which would 
become possible at  HERA,  will give more information to help separate 
the models and help us understand the phenomena which 
limit the rise of the nucleon structure function $F_2$  
at small $x$; e.g. see references \cite{KMS,MQ}.  

\noindent
\begin {bf} Is it feasible to study shadowing in HERA? \end {bf}

To answer this question we assume that nuclear beams can be stored in HERA 
and luminosities of 2 pb$^{-1}$ per nucleon, shared between 2 nuclear targets, 
can be achieved (i.e. $2/A$~pb$^{-1}$ 
per nucleus, where $A$ is the atomic weight).  With this definition 
of luminosity, rate computations should use cross sections per nucleon.   
The counting rates in bins of $Q^2$  and $x$  are then estimated to assess 
the statistical errors on the measurements of the ratios of 
$F_2^{A}/F_2^D $ where the nuclear targets are assumed to be
He, C or S.  The cross sections are calculated from the one 
photon exchange formula

\begin{equation}
\frac{d^2\sigma}{dxdQ^2} = 
 \frac{4\pi\alpha^2}{xQ^4}\left[ 1 - y + \frac{y^2}{2(1+R)}\right]
F_2(x,Q^2).
 \end{equation}

\noindent
The nucleon structure functions, $F_2$ and $R$, are computed from the 
MRS(A) set \cite{MRSA}.  Deep inelastic scattering events are assumed to 
be detectable with 100\% efficiency if the scattered electron energy 
$E^\prime >5$ GeV and its scattering angle is more than 3 degrees to 
the electron beam.  Alternatively events are assumed to be detectable 
with 100\% efficiency if a quark is scattered out by an angle of more 
than 10 degrees to the proton beam and with an energy more than 5 GeV.  
Nuclear effects on the structure functions are neglected.  Such effects, 
which are at the 10\% level, will have little effect on the statistical 
accuracy of measurements of ratios of structure functions, $F_2^A/F_2^D$.  

\begin{figure}[hbt] 
\begin{center}
\vspace{2mm}
\epsfig{file=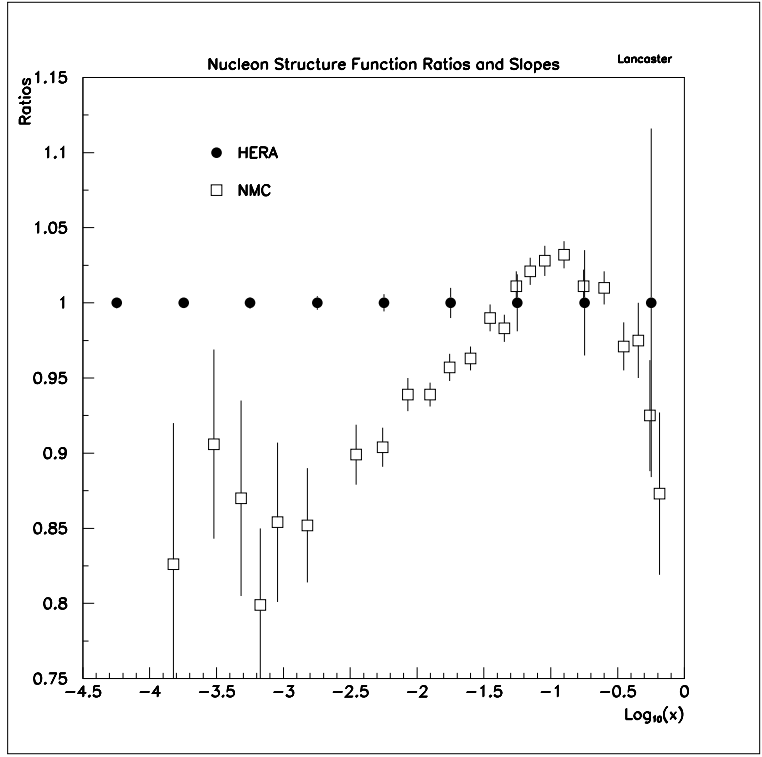,width=10cm}
\caption[junk]{ Ratio 
of the nucleon structure function in 
carbon to that in
 deuterium as a function of $x$.  The NMC data \cite{nmc_li} (open squares) are
 shown in comparison to data with the estimated statistical accuracy
 of an experiment of luminosity 1 pb$^{-1}$ per nucleon at HERA. }
\label{shadac1}
\end{center}
\end{figure}

Fig.~\ref{shadac1} shows the estimated statistical errors on the 
ratios (averaged over $Q^2$),  with these assumptions, together with the 
measurements of the NMC~\cite{nmc_li}.  In many cases the statistical errors 
are $ < {1\%}$ i.e. smaller than the sizes of the points.  
It can be seen from Fig.~\ref{shadac1}  that this statistical precision 
will allow high accuracy measurements of the shadowing ratios down to 
lower $x$ values than in fixed target experiments and over a much wider 
range of $Q^2$.  

\begin{figure}[htb] 
\begin{center}
\epsfxsize10cm\leavevmode\epsfbox{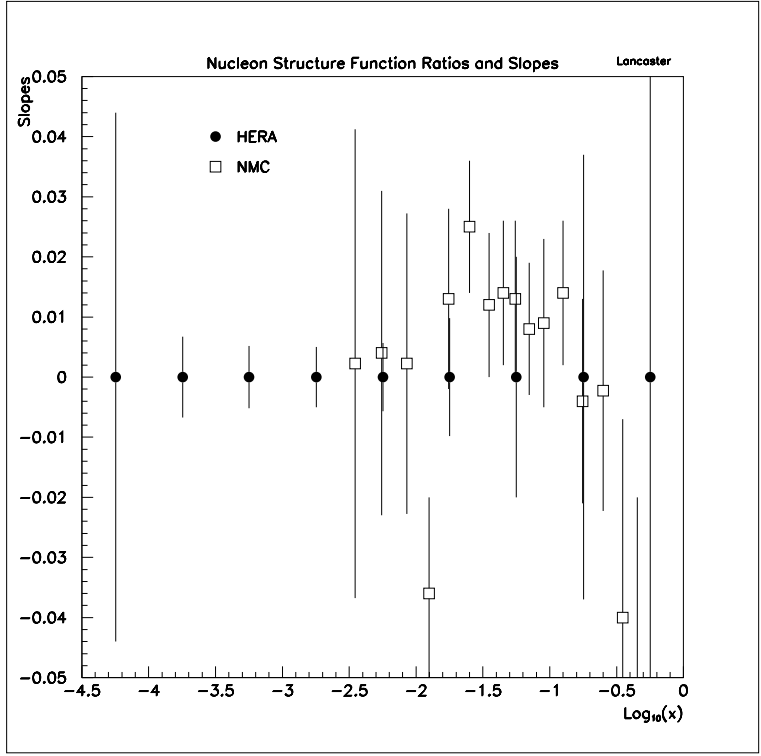}
\caption[junk]{
The slopes $d(F_2^A/F_2^D)/d \ln{Q^2}$ as 
function of $x$ showing the NMC data~\cite{nmc_rean}
 (open squares) and the statistical accuracy of an experiment with 1 pb$^{-1}$
 per nucleon at HERA.  }
\label{shadac2}
\end{center}
\end{figure}

Fig.~\ref{shadac2} shows the statistical precision of the slopes 
$d(F_2^{A}/F_2^D)/d\ln{Q^2}$ estimated at HERA compared to the NMC data.  
Impressive precision is  possible at HERA, presumably due to the much 
larger $Q^2$ range covered.  Measurements over such a large $Q^2$ range 
will allow the precise predictions of the parton fusion model to be tested.  
 
If at least two different nuclear targets can be stored in different 
bunches in HERA during experimental running the systematic errors should 
be similar in magnitude to those in fixed target experiments 
\cite{nmc_rean}.  
Radiative corrections for deep inelastic scattering from heavy nuclear 
targets will be necessary and these will be applied with the appropriate cuts 
on the data as discussed in these preceedings \cite{Kurek,AANDS}.  

 In conclusion, measurements of the ratios $(F_2^{A}/F_2^D)$ at 
HERA will extend the data to smaller values of $x$ and much larger values 
of $Q^2$ than in fixed target experiments. Impressive precision on the 
$Q^2$ dependence will be possible from which the mechanism which leads to the 
limitation in the rise of $F_2$ at small $x$ and large $Q^2$ can be studied.

\subsection{The Accuracy of The Gluon Density Measurements Using Nuclei in HERA}
\subsubsection{Introduction}

The major contribution to shadowing from nonlinear QCD effects is thought 
to arise from multigluon interactions such as gluon fusion effects.  
Such effects are amplified at higher 
$x$ in nuclei due to the larger target size so that they should become 
visible in the HERA kinematic range.  Similar effects are 
expected to limit the growth of the nucleon structure function $F_2$ at 
high $Q^2$ and low $x$.  Hence it is interesting to look for such 
effects directly on the gluon distribution by looking for differences 
between the gluon density in bound and free nucleons.  
Different ways of doing this are studied below.  

\subsubsection{Determination of the Gluon Density in the Nucleus from the 
Jet Rates}

\begin{figure}
\begin{center}
\leavevmode
\hbox{%
\hspace*{-0.8cm}
\epsfxsize = 15cm
\epsffile{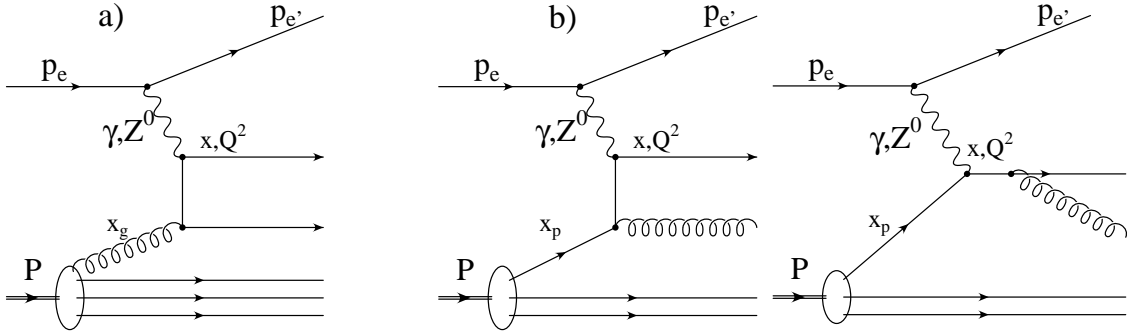}}
\end{center}
\vspace{-0.5cm}
\caption{The processes giving rise to $2+1$ jet topology.
P denotes here the projectile particle, a proton or a nucleus, $Q^2$
is the four-momentum transfer and $x$ is the Bjorken variable.  
}
\label{diagrams}
\end{figure}

In the majority of large $Q^2$ deep inelastic electron-nucleus scattering 
events, at HERA, the scattered quark and the remnant of the nucleus
will hadronize to form two jets.
Such a topology is  often called the  $1+1$ jet configuration.
Occasionally, however, more jets will be produced.  In Fig.~\ref{diagrams} 
the partonic processes giving rise to the $2+1$ jet topology
are shown. These processes are called ``boson-gluon fusion'' (process a)
and ``the QCD Compton scattering" (processes b).

The $2+1$ jet events can be related to the partonic processes shown 
in Fig.~\ref{diagrams} if the invariant mass of the system of 
two jets $ \overline{s} = (p_{jet1}+p_{jet2})^2$ is significantly larger 
than the typical scale of the strong interactions, so that perturbative QCD
can be used.  The contribution of these
processes to the total cross section depends upon the value of the 
coupling constant $\alpha_s$ defining the strength of quark-gluon coupling 
and upon the momentum distribution of the incoming  gluon (quark).  
The fractions of the parent bound nucleon momentum carried by 
the incoming partons, $x_{p,g}$, are constrained by the value of $x$, 
the total hadronic mass, $W$, 
and the invariant mass of the two jet system $\overline{s}$:
\begin{eqnarray}
x_{p,g}= x + \overline{s}/W^2.
\end{eqnarray} 
For values of $\overline{s}/W^2 \geq 0.01$, 
the values of $x_{p,g}$ must be large. In this kinematic domain the 
partonic distributions in bound nucleons have been well measured in 
fixed target experiments~\cite{reviews}.  Thus, the coupling constant, 
$\alpha_s$, can be derived from the measured rate of $2+1$ jet events.  
In turn, we shall be able to use this $\alpha_s$ value to determine 
the gluon momentum distribution at smaller $x_{g}$, corresponding to 
small values of both $x$ and  $\overline{s}/W^2$ (note that at small $x$ the 
contribution of the QCD Compton processes (Fig.~\ref{diagrams}b,c) 
to the total ``2+1" jet cross section 
is expected to be small). Such an analysis has been recently carried out by the  
H1 collaboration \cite{H1GLUON} using deep inelastic electron-proton 
scattering data collected at HERA. The systematic errors of the resulting 
gluon density are large and to some extent uncertain.  They are dominated 
by the uncertainties in relating the measured rate of observed jet topologies 
to the basic QCD processes involving quarks and gluons.  
Unfolding the gluon densities involves modelling the hadronisation of 
the quarks which is necessary for simulating the detector effects 
but is only weakly constrained by the data. In addition several jet 
algorithms can be used leading to differences in jet counting.  
There exist as well ambiguities in the QCD calculation of the 
processes shown in Fig.~\ref{diagrams}.  
The amplitudes of the processes $\gamma^{*}g \rightarrow jet_1 + jet_2$ 
and  $\gamma^{*}q \rightarrow jet_1 + jet_2$ have poles corresponding 
to the collinear emission of jets with respect to the direction of the 
incoming gluon ($\gamma^{*}$)  and to the emission of jets of small invariant 
mass.  Since one must use fixed order perturbative QCD,  
 jets reconstructed in phase space close to the poles must be 
 avoided to diminish the sensitivity to  higher order terms which have so far 
not been calculated. This can be achieved by using only jets 
 of high invariant mass 
 $ (\overline{s} \geq 100$~GeV$^2$) and by rejecting events in which 
 one of the jets is emitted at a small angle with respect to the incoming 
 proton direction.  
  In the region of small  $\overline{s}$ the jet rates calculated using 
leading order and next to leading order approximations 
 are significantly different \cite{graudenz} indicating that higher order 
 corrections are necessary for an unambigous determination of the gluon 
density.  

\begin{figure}
\vspace{-1.5cm}
\begin{center}
\leavevmode
\epsfig{file=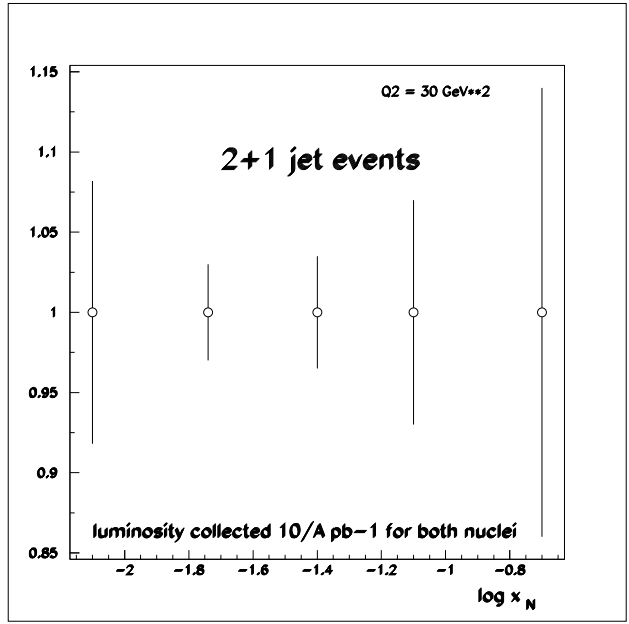,height=8cm,bbllx=25 ,bblly=148 , bburx=566 , bbury=689}
\end{center}
\vspace{-0.3cm}
\caption{
Statistical precision of the ratio of gluon densities
in two nuclei $x_N G(x_N,A_1)/xG(x_N,A_2)$  at $Q^2=30$~GeV$^2$ determined
from the $``2+1"$ jet sample.
}
\label{gluon2}
\end{figure}

Will it be possible to use the jet method to determine the gluon density 
in nuclei given the uncertainties described above and at the same time  
adding the extra ambiguity related to jet formation in the nuclear 
medium?  Most likely it will be difficult to obtain satisfactory precision in 
measuring the absolute gluon distributions in the nuclei. 
However,  we expect that good precision can be achieved in measuring the 
ratios of gluon distributions for various nuclei.  
The uncertainties due to the jet finding algorithms and to the modelling of 
the hadronisation processes will largely cancel in the ratios if 
high energy jets are used.  
High energy jets are expected to be formed outside the nucleus. In addition we
shall be able to select jets in the restricted phase space region where the
effects of rescattering of slow particles belonging to the jet but formed 
inside the nucleus are small. The energy loss of quarks and gluons
traversing the nuclear medium prior to hadronisation is expected to
be below 350 MeV/fm according to the estimation of \cite{Baier}
and should not give rise to large errors on the gluon density 
ratio.  The expected statistical precision of the measurements of the ratio 
of the gluon densities 
for two nuclei of atomic numbers A1 and A2 is shown in Fig.~\ref{gluon2}.
It corresponds to a luminosity of 10 pb$^{-1}/A$ for each nucleus.  
In estimating this statistical precision we have followed the jet 
and kinematic region selection of \cite{H1GLUON} and neglected all nuclear 
effects.  We also assume that the contribution of the QCD Compton process 
can be unambigously subtracted.  The $x_N$ variable is the fraction of the 
bound nucleon momentum carried by the gluon.  It is clear that good 
statistical precision on the measurement of the ratio of the gluon densities
can be achieved at modest luminosities.  

\subsubsection{The Gluon Distribution in Nuclei from Scaling Violations} 

The gluon distribution can also be determined from the deviations from scaling 
of the structure function $F_2$.  We estimate 
the accuracy of such a determination of the gluon density for bound nucleons 
in an experiment with nuclear beams in HERA.  
The scaling violations of $F_2$ are strongly 
related to the gluon distribution of the target nucleon at small $x$ values.  
The quantity 
\begin{equation}
 \psi = \frac{\frac{dF_2^{A}}{d \ln Q^2} - \frac{dF_2^D}{d \ln Q^2}} 
{\frac{d F_2^p}{d \ln Q^2}}
 \end{equation}
\noindent
is sensitive to the differences of the gluon distribution in bound and free
nucleons and is roughly proportional to $\delta G$/$G$ where $G$ is the gluon
density at a particular $x$ value and $\delta G$ is the difference in gluon
densities between bound and free nucleons.  Hence it would be interesting to
measure this quantity in an experiment with nuclear beams stored in HERA.  

   To estimate the feasibility of such a measurement the accuracies of the
determinations of the slopes $dF_2/d\ln{Q^2}$ have been estimated from 
the expected counting rates for a 2 pb$^{-1}$ per nucleon run using 
the MRS(A) set of structure functions \cite{MRSA}.  The slopes were obtained 
from a linear least squares fit to values of $F_2$ using the statistical errors 
calculated for such an experiment with the assumptions described in 
section~\ref{The Accuracy of Shadowing Measurements Using Nuclei in HERA}.  
The error in the quantity $\psi$ was 
then obtained assuming equal statistical errors for the nuclear and deuteron 
targets with a 1 pb$^{-1}$ per nucleon run for each.  The error on the slope 
$dF_2^p/d\ln{Q^2}$ for the proton was neglected since this should be well 
determined from high luminosity proton running.  

\begin{figure}
\begin{center}
\vspace{-1.5cm}
\epsfxsize10cm\leavevmode\epsfbox{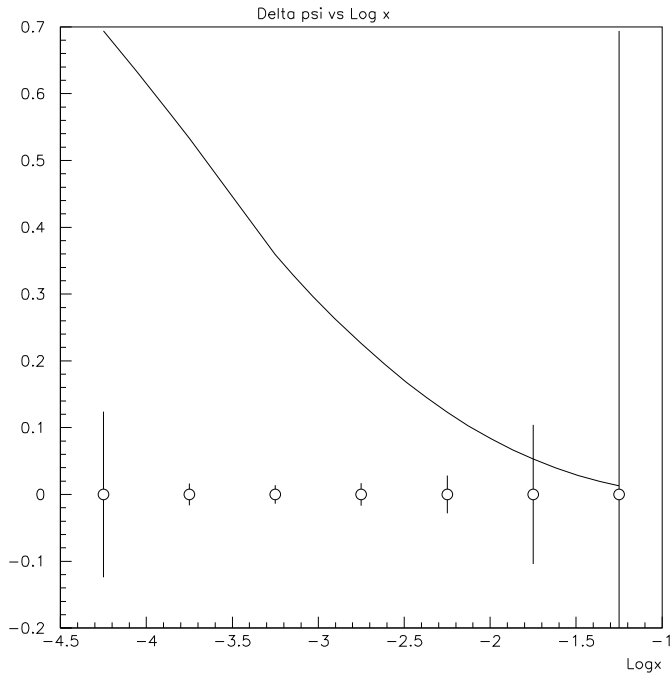}
\end{center}
\vspace{-0.9cm}
\caption{
Accuracy of the difference between the slopes of $F_2$ between
nuclei and deuterium as a function of $x$.  The smooth curve shows the 
calculated slope for the free proton. }
\label{HESLGLU}
\end{figure}

   The results are shown in Fig.~\ref{HESLGLU}.  For the purpose of this
figure  we assume that the gluon densities for bound and free nucleons
are the same so that $\psi$=0.  The estimated statistical errors are then
superimposed on these values and the smooth curve shows 
$dF_2^p/d\ln Q^2$ for comparison computed from the MRS(A) set.  
Comparison of the errors on $\psi$ with the values of the slope for the 
proton shows that this method will be sensitive to differences in the 
gluon densities between bound and free nucleons of the order of 5 per cent 
of the total gluon density at $x$ in the vicinity of 10$^{-3}$.  Hence the 
measurement will be quite sensitive to nuclear effects on the gluon density.

\subsubsection{Determination of the Gluon Density in the Nucleus 
from Inelastic $J/\psi$ production}

The expected accuracy of the determination of the bound 
to free nucleon gluon density ratio 
$xG|_A/xG|_D$ using nuclei in HERA is estimated for 
inelastic photoproduction of $J/\psi$ mesons at $Q^2<4$~GeV$^2$
($eA \rightarrow eXJ/\psi$).

Inelastic production of $J/\psi$ mesons has been used for a long time to 
extract or constrain the gluon distribution in the nucleon (see 
e.g.~\cite{emc_psi}-\cite{h1warsaw}), assuming that the dominant mechanism is 
the photon-gluon fusion~\cite{weiler}-\cite{martin-ng}.
Within this framework, the cross section 
is directly proportional to the gluon density.
These calculations have been affected in the past by large 
normalisation uncertainties (up to factors of 2-5), which however 
have been greatly reduced recently~\cite{kraemer}. 
Inelastic $J/\psi$ production dominates at values of $z$, the fraction 
of the photon energy carried by the meson in the nucleon rest frame, 
smaller than $\approx 0.9$.
In the framework of the colour singlet model~\cite{martin-ng}, 
the gluon distribution is probed at a value of $x$, the fraction of
the proton's momentum carried by the gluon,
$x =[m_{J/\psi}^2/z+p_t^2/z(1-z)]/W^2$, where $p_t$ is the $J/\psi$
transverse momentum with respect to the virtual photon direction and $W$ is the
photon-nucleon centre-of-mass energy. The scale probed by this 
process is approximately $m_{J/\psi}^2 \approx 10$~GeV$^2$.

\begin{figure}[ht]
\vspace{-1.0cm}
\begin{center}
\leavevmode
\hbox{%
\hspace*{-0.8cm}
\epsfxsize = 10cm
\epsffile{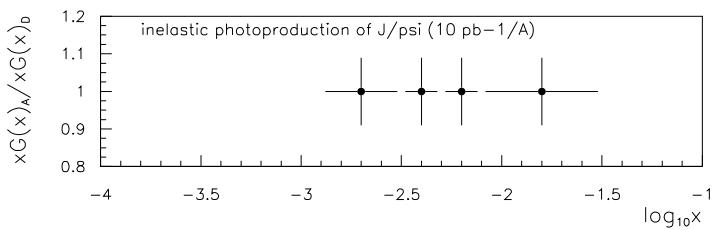}}
\end{center}
\vspace{-7.0cm}
\caption{Expected statistical accuracy (vertical error bars) on 
the ratio $xG|_A/xG|_D$ as a function of $\log_{10}{x}$ using inelastic 
photoproduction ($Q^2<4$~GeV$^2$) of $J/\psi$ mesons. An integrated 
luminosity of 10~pb$^{-1}/A$ for each nucleus has been
assumed. The horizontal bars indicate the size of the bins.
}
\label{accuracy_psi_in}
\end{figure}

Fig.~\ref{accuracy_psi_in} shows the expected statistical accuracy as
a function of $\log_{10}{x}$ for an integrated luminosity of 
10~pb$^{-1}/A$.
Decays into $e^+e^-$ or $\mu^+ \mu^-$ pairs have been assumed. The 
plot refers to the kinematic region $Q^2<4$~GeV$^2$, $z<0.9$.
Nuclear effects have been neglected in the evaluation of statistical accuracies.

Possible sources of systematic uncertainties are the luminosity, the
branching ratio, the global acceptance (including trigger 
and reconstruction efficiency, muon or electron identification etc.),
the feed-in from $\psi^{\prime}$ production and the contamination
from resolved photon events. In the recent 
H1~\cite{h1_el_inel} and ZEUS~\cite{zeuswarsaw} analyses, 
the total systematic uncertainty is approximately 10-20\%. 
By and large all the above contributions would cancel in a ratio 
for simultaneously stored nuclei, with the partial exception of the luminosity.
In practice, for an integrated luminosity of 10~pb$^{-1}$, the statistical
uncertainty dominates.

\subsection{Diffraction from Nuclei in HERA}
\subsubsection{Introduction}

The measurements of the ZEUS and H1 collaborations \cite{Zrapgap,H1rapgap}
of deep-inelastic electron-proton scattering have revealed
the existence of a distinct class of events
in which there is no hadronic energy flow in an interval of
pseudo-rapidity, $\eta$, adjacent to the proton beam
direction i.e.  events with a large rapidity gap.  
Such events are interpreted as deep inelastic scattering from the pomeron, 
$I\!\!P$.  Studies of events with a large rapidity gap from nuclear 
targets will allow the structure of the pomeron from a different source 
than the free nucleon to be determined.  It will be interesting to see if 
these structures differ.  In addition, the study of diffractive vector meson 
production will be interesting to search for the phenomenon of colour 
transparency.  Such a phenomenon has not yet been convincingly seen 
although it is predicted in QCD.  

\subsubsection{Expected Accuracy of the Measurements of the Pomeron Structure}

\begin{figure}
\vspace{-1.5cm}
\begin{center}
\leavevmode
\hbox{%
\hspace*{-0.8cm}
\epsfxsize = 8cm
\epsffile{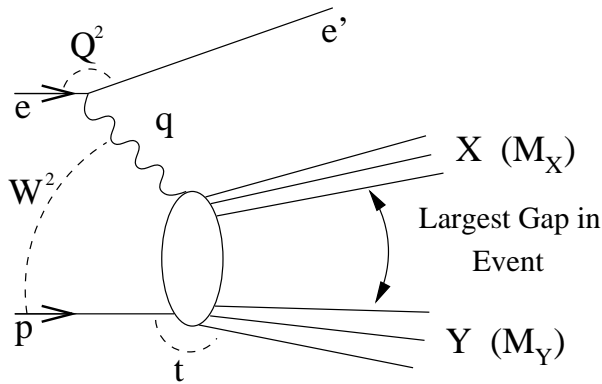}}
\end{center}
\vspace{-0.5cm}
\caption{
The diagram of the process with a rapidity gap
between the system X and Y. The projectile nucleus  is denoted here as p.}
\label{diff1}
\end{figure}

We commence by describing the terminology surrounding measurements of the 
Pomeron structure.  
In our studies we shall use the four variables
$\beta_A$, $Q^2$, $x_A$ and $t_A$, or equivalently $\beta_A$, $Q^2$,
$x_{I\!\!P,A}$ and $t_A$, which are defined as follows:
\begin{equation}
x_A=\frac {-q^2}{2P\cdot q}; \,\,\,\,\,\,
x_{I\!\!P,A} = \frac{q\cdot (P-P')}{q\cdot P}; \,\,\,\,\,\,
Q^2 = -q^2; \,\,\,\,\,\,
\beta _A = \frac{-q^2}{2q\cdot (P-P')}; \,\,\,\,\,\,
t_A=(P-P')^2.
\label{eq:definition}
\end{equation}
Here $q$, $P$ and $P'$ are, as indicated in Fig.~\ref{diff1},  the $4$--momenta
of the virtual boson, incident
nucleus  and the final state colourless remnant Y respectively.
The latter can be
either a coherently recoiling nucleus or
any incoherent excitation of the nucleus carrying its quantum numbers.
The variables in equation (\ref{eq:definition}) are related to each other 
via the expression: 
\begin{equation}
x_A=\beta _A x_{I\!\!P,A}.
           \label{eq:trivial}
\end{equation}
We also introduce, in order to allow the  comparison  of measurements with
 nuclei of different atomic number  $A$,
the variables:
\begin{equation}
x=x_A \cdot A \,\,\,\,\,\,\,\,
\,\,\,\,\,\,\,
x_{I\!\!P} = x_{I\!\!P,A} \cdot A \,\,\,\,\,\,\,\,\,
\,\,\,\,\,\,
\beta  = \beta _A \,\,\,\,\,\,\,\,\,\,\,\,\,\,\,
t=t_A.
           \label{eq:definition1}
\end{equation}
Note that relation (\ref{eq:trivial}) rewritten in terms of the above
variables still holds i.e.
\begin{equation}
x=\beta  x_{I\!\!P}.
           \label{eq:trivial1}
\end{equation}
The $A$-rescaled variables can be directly related to the variables defined in
\cite{H1F3D,ZEUSF3D} retaining their interpretation, as was given there, 
for processes in which only  one nucleon of the nucleus interacts with 
the electron and the nucleon's  Fermi momentum can be  neglected.

The variables $x_{I\!\!P,A}$, $x_{I\!\!P}$ and $\beta$
can be expressed in terms of the
invariant mass of the hadronic system $X$, $M_X$, the nucleus mass, $M_A$, and
the total hadronic
invariant mass $W$ as
\begin{equation}
x_{I\!\!P,A}  = \frac{Q^2+M_X^2-t}{Q^2+W^2-M_A^2}
        \approx \frac{Q^2+M_X^2}{Q^2}\cdot x_A,
                   \label{eq:xpomQ2}
\end{equation}
\begin{equation}
x_{I\!\!P}    \approx \frac{Q^2+M_X^2}{Q^2}\cdot x,
                   \label{eq:xpom1Q2}
\end{equation}

\begin{equation}
\beta = \beta_A = \frac{Q^2}{Q^2+M_X^2-t} \approx \frac{Q^2}{Q^2+M_X^2}.
           \label{eq:beta}
\end{equation}

In the kinematic domain which we shall consider here
($M_A^2\ll W^2$
and $\mid\!t\!\mid\ll Q^2$, $\mid\!t\!\mid\ll M_X^2$)
$x_{I\!\!P,A}$ may be interpreted as the fraction of the
$4$--momentum of the nucleus carried by the $I\!\!P$
and $\beta$ as the fraction of
the $4$--momentum  of the $I\!\!P$ carried by the quark interacting
with the virtual boson. Note that in the interactions in which only one
nucleon takes part,  $x_{I\!\!P}$ can be interpreted as the fraction of the
$4$--momentum of this nucleon carried by the  $I\!\!P$.

We shall follow in our analysis the formalism defined in 
\cite{H1F3D,ZEUSF3D} and introduce
the diffractive structure function $F_{2,A}^{D(3)}$ as a function of three
kinematic variables, derived from the structure function $F_{2,A}^{D(4)}$.
This latter structure function depends upon
four kinematic variables chosen here as: $x$, $Q^2$, $x_{I\!\!P}$ and $t$ and
is defined by analogy with the decomposition of the
unpolarised total $eA$ cross section. The total cross section can be expressed
in terms of two
structure functions $F_{2,A}^{D(4)}$ and
$\frac{F_{2,A}^{D(4)}}{2x(1+R_A^{D(4)})}$
in the form
\begin{equation}
\frac{{\rm d}^4 \sigma_{eA\rightarrow eXY}}{{\rm d}x{\rm d}Q^2{\rm
d}x_{I\!\!P}dt}
= A \cdot \frac{4\pi \alpha^2}{xQ^4}\,
  \left\{1-y+\frac{y^2}{2[1+R_A^{D(4)}(\beta ,Q^2,x_{I\!\!P},t)]}\right\}
  \,F_{2,A}^{D(4)}(\beta ,Q^2,x_{I\!\!P},t),
           \label{eq:defF2D}
\end{equation}
in which $y=Q^2/sx_A$ and  $s$ is the
$eA$ collision centre of mass (CM) energy squared.
We shall discuss here the low $y$ region and neglect the term containing
$R_A^{D(4)}$.

We shall consider in the following measurements of
$\frac{d^3 \sigma (eA\rightarrow eXY)}{{\rm d}x{\rm d}Q^2{\rm d}x_{I\!\!P}}$,
from which the structure function
$F_{2,A}^{D(3)}(\beta ,Q^2,x_{I\!\!P})=\int F_{2,A}^{D(4)}(\beta
,Q^2,x_{I\!\!P},t)\,{\rm d}t$
can be derived.
The integration is over the range
$\mid\!t_{min}\!\mid<\mid\!t\!\mid<\mid\!t\!\mid_{lim}$ where $t_{min}$ is a
function of $Q^2$, $W^2$, $\beta$ and the mass of the
the system Y, and $\mid\!t\!\mid_{lim}$ is specified by the
requirement that all particles belonging to the system Y remain undetected.
The structure function
$F_{2,A}^{D(3)}$ will thus be derived  from
\begin{equation}
\frac{{\rm d}^3 \sigma_{eA\rightarrow eXY}}{{\rm d}x{\rm d}Q^2{\rm d}x_{I\!\!P}}
= A \cdot \frac{4\pi \alpha^2}{xQ^4}\,
  \left\{1-y+\frac{y^2}{2}\right\}
  \,F_{2,A}^{D(3)}(\beta ,Q^2,x_{I\!\!P}).
           \label{eq:F2D3}
\end{equation}

 The structure function
 $F_{2,A}^{D(3)}(\beta , Q^2,x_{I\!\!P})$
 can  be measured at HERA using a sample of ``rapidity gap" events i.e. events in
which
 there is no hadronic energy flow over  a large  $\eta$ interval.
 These events originate  from coherent diffractive scattering
 ($eA \rightarrow eA + X_{diff}$) and from incoherent diffractive scattering
 ($eA \rightarrow e(A-N) + N  + X_{diff}$). $N$ denotes here the nucleons
ejected
 from the incoming nucleus.
 The measurement of $F_{2,A}^{D(3)}(\beta ,Q^2,x_{I\!\!P})$ for several nuclei,
separately
 for coherent and incoherent processes \cite{KRASNY}, will provide an 
important test of  the
 quark parton interpretation of diffractive processes and a unique means to
 find out how universal is the concept of the pomeron.

\begin{figure}
\vspace{-1.7cm}
\begin{center}
\leavevmode
\epsfig{file=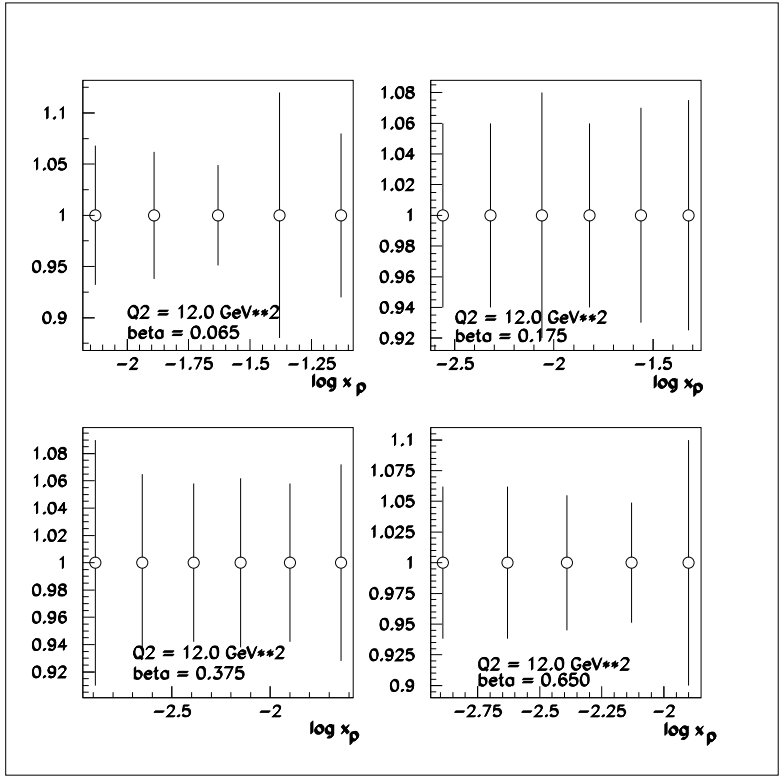,height=10cm,bbllx=25 ,bblly=148 , bburx=566 ,bbury=689 }
\end{center}
\vspace{-0.2cm}
\caption{
The ratio $R_{A1,A2}(\beta ,Q^2,x_{I\!\!P})$
plotted as a function of $x_{I\!\!P}$ for fixed values of $\beta$ 
and  $Q^2$. The error bars correspond to a luminosity of
$10/A1$ and $10/A2$ pb$^{-1}$ for each nucleus.  }
\label{diff2}
\end{figure}

  We propose  to measure the ratio of the structure
 functions
 \begin{equation}
 R_{A1,A2}(\beta ,Q^2,x_{I\!\!P}) = F_{2,A1}^{D(3)}(\beta ,Q^2,x_{I\!\!P}) /
  F_{2,A2}^{D(3)}(\beta ,Q^2,x_{I\!\!P}),
 \end{equation}
 where A1 and A2 denote the atomic numbers of the two nuclei.
 This ratio can be measured at HERA with a very high systematic accuracy.
 The statistical precision
 of such a measurement will be  of the order of 5 \%, if luminosities  of
 $10/A1$ and $10/A2$~pb$^{-1}$ are  collected for each nucleus.
 This is illustrated in  Fig.~\ref{diff2} where we show, as an example,
 $R_{A1,A2}$  at $Q^2= 12$ GeV$^2$  as a function of $\beta$ and $x_{I\!\!P}$.
 In order to estimate the statistical precision of the measurement we have used
the
 RAPGAP \cite{jung} Monte-Carlo and have assumed the event selection procedure
of \cite{H1F3D,ZEUSF3D}.  
 We have not tried to model the nuclear dependence of the
$F_{2,A2}^{D(3)}(x,Q^2,x_{I\!\!P})$
 resulting in  the ratio shown in Fig.~\ref{diff2} to be  equal 1.
 
Several distinct hypotheses concerning the deep inelastic structure of the
diffractive
processes can be verified  (rejected) by measuring the $R_{A1,A2}$:
\begin{itemize}
\item
universal (independent of the source) pomeron structure and an $A$-independent
pomeron flux leading to
\begin{eqnarray}
R_{A1,A2}(\beta ,Q^2,x_{I\!\!P})=1;
\end{eqnarray}
\item
universal pomeron structure and an $A$-dependent pomeron flux leading to
\begin{eqnarray}
R_{A1,A2}(\beta ,Q^2,x_{I\!\!P})=f(A1,A2);
\end{eqnarray}
\item
$A$-independent pomeron flux and a parent nucleus dependent pomeron structure.
In a model of this type \cite{Buchmuller} the ratio $R_{A1,A2}$ can be expressed
using the
nuclear structure functions $F_{2,A}(x,Q^2) $
measured in inclusive electron nucleus scattering
\begin{eqnarray}
 R_{A1,A2}(\beta ,Q^2,x_{I\!\!P})= F_{2,A1}(\beta \cdot
x_{I\!\!P},Q^2)/F_{2,A2}(\beta \cdot x_{I\!\!P},Q^2).
\end{eqnarray}
\end{itemize}

\subsubsection{Measurement of the $A$-dependence of the Fraction 
of Rapidity Gap Events}

\begin{figure}
\vspace{-1.2cm}
\begin{center}
\leavevmode
\epsfig{file=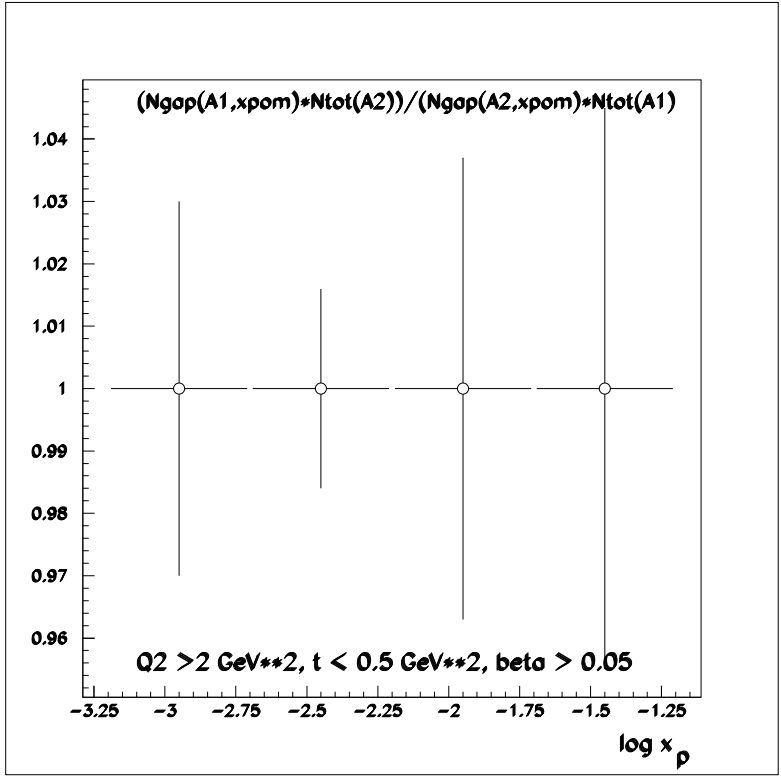,height=10cm,bbllx=25 ,bblly=148 , bburx=566 ,bbury=689 }
\end{center}
\vspace{-0.5cm}
\caption{The integrated ratio $R_{gap}(A1,A2)$
plotted as a function of $x_{I\!\!P}$. The
error bars correspond to a luminosity of
$1/A1$ and $1/A2$~pb$^{-1}$  for each nucleus.  }
\label{diff3}
\end{figure}

One of the simplest measurements which could discriminate between the 
two pictures of pomeron formation proposed in \cite{Buchmuller,FSAGK} 
and unresolved by the diffractive $ep$ scattering data,
is the measurement of the $A$-dependence of the fraction of the number 
of rapidity gap events with respect to the total number of deep inelastic 
scattering events:  
\begin{eqnarray}
R_{gap}(A1,A2) = \frac{N_{gap}(A1)/N_{tot}(A1)} {N_{gap}(A2)/N_{tot}(A2)}.
\end{eqnarray}
In the model of \cite{Buchmuller} the pomeron is replaced by  multiple
soft colour exchanges between the quark-antiquark pair
into which the virtual photon has fluctuated and the target nucleus.
In this model the ratio $R_{gap}(A1,A2)$ is expected to be 1.
This is in contrast to the prediction of the colour
singlet exchange model \cite{FSAGK} in which the ratio $R_{gap}(A1,A2)$
can reach a value of 3 between  nuclei with  atomic numbers A1 and A2 
differing by 200.
In order to illustrate the statistical precision
of such a measurement we show in Fig.~\ref{diff3} the values of
$R_{gap}(A1,A2)$ and their statistical errors.

The statistical precision of $R_{gap}(A1,A2)$
is dominated by the statistical precision of  $N_{gap}(A1)$  and  $N_{gap}(A2)$
and was  determined with help of the
 RAPGAP \cite{jung} Monte Carlo by counting  events generated within  the
kinematic
 domain defined by the following boundaries: $Q^2 \geq 2$ GeV$^2$, $ |t|\leq
0.5$ GeV$^2$ and
 $\beta \geq 0.05$.
The ratio shown in Fig.~\ref{diff3} was set to unity  as
we have not tried to model the nuclear dependence of the $R_{gap}(A1,A2)$.
We observe that
statistical precisions of $\simeq 2 ~\%$ can be reached by collecting
integrated luminosity of $1/A$ pb$^{-1}$ for each nucleus.  The systematic errors,
as for the measurement  of $F_{2,A}^{D(3)}(\beta ,Q^2,x_{I\!\!P})$
 are expected to be smaller than the statistical
errors if two nuclei are stored simultaneously at HERA.
This measurement, even if carried out for two light nuclei, can easily rule out
one of the
two models of pomeron formation.

\subsubsection{Elastic Vector Meson Production}

We present the accuracy expected for 
exclusive photoproduction of $J/\psi$ mesons at $Q^2<4$~GeV$^2$
($eA \rightarrow eAJ/\psi$) and
exclusive production of $\rho^0$ mesons at $Q^2>4$~GeV$^2$ 
($eA \rightarrow eA \rho^0$).

As discussed in detail later, elastic (or exclusive) production of 
vector mesons in the reaction $ep \rightarrow ep V$, where $V$ is 
a vector meson ($\rho^0$, $\omega$, $\phi$, $J/\psi$...), 
is thought to be of a diffractive nature.  Recent calculations 
(cf. e.g.~\cite{misha,Brod94,nemchik,FKS,misha2}) indicate that 
the cross section for these processes may depend on 
the gluon momentum distribution $\bar x G(\bar x)$ probed at a value of 
$\bar x$, the fraction of the nucleon's momentum carried by the 
gluon, $\bar x \simeq (Q^2 +m_V^2)/W^2$, where $m_V$ is the vector meson 
mass 
and $W$ is the photon-nucleon centre-of-mass energy.  

\begin{figure}[ht]
\vspace{-1.0cm}
\begin{center}
\vspace{-0.3cm}
\leavevmode
\hbox{%
\hspace*{-0.8cm}
\epsfxsize = 10cm
\epsffile{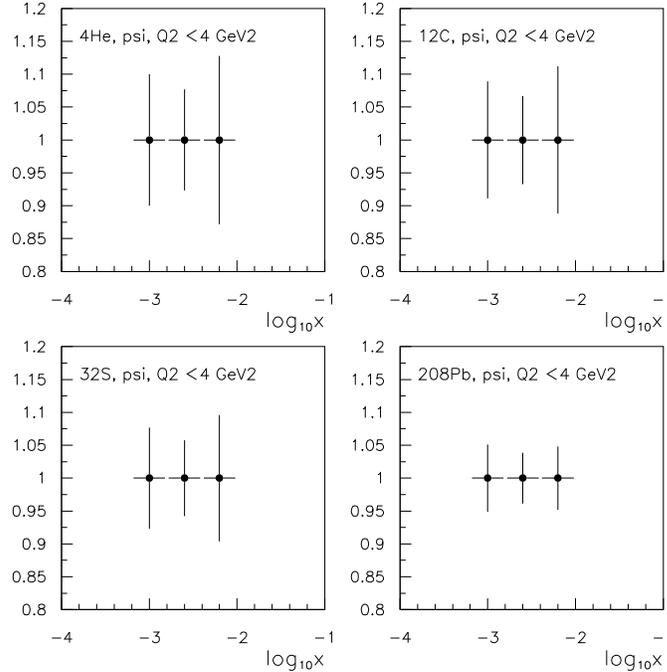}}
\end{center}
\vspace{-1cm}
\caption{Expected statistical accuracy (vertical error bars) on 
the ratio $R{_V^{1/2}}$ for 
elastic photoproduction ($Q^2<4$~GeV$^2$) of $J/\psi$ mesons.
In all cases a luminosity of 1~pb$^{-1}/A$ for each nucleus was 
assumed. The horizontal bars indicate the size of the bins.}
\label{psi_el}
\end{figure}

We have determined the statistical accuracy with which the ratios 

\begin{equation}
R_V=
\frac{1}{A^2}\frac{d\sigma^A/dt|_{t=0}}{d\sigma^D/dt|_{t=0}}
 \end{equation}

\noindent 
can be determined for $^4$He, $^{12}$C, $^{32}$S and $^{208}$Pb. 
Fig.~\ref{psi_el} shows the results for $J/\psi$ production.

We used DIPSI~\cite{dipsi}, a Monte Carlo
generator based on~\cite{misha} that describes the available ZEUS data
on $J/\psi$ photoproduction.
Events were generated in the range $Q^2<4$~GeV$^2$ and
$30<W<300$~GeV; the $J/\psi$ was assumed to decay into 
$e^+e^-$ or $\mu^+ \mu^-$ pairs. Events in which the decay leptons
were outside the coverage of the barrel and rear tracking detectors
of H1 and ZEUS (approximately $34^{\circ} < \vartheta < 164^{\circ}$)
were rejected. A further reduction in the number of accepted events 
by a factor 0.2 was applied to account for efficiency and acceptance 
effects; this factor was taken to be independent of $\bar x$.

Only statistical uncertainties are shown. 
Possible sources of systematic uncertainties are the luminosity, the
branching ratio, the global acceptance (including trigger 
and reconstruction efficiency, muon or electron identification etc.),
the contamination from incoherent events,
the feed-in from inelastic $J/\psi$ production (\`a la photon-gluon fusion)
and the feed-in from $\psi^{\prime}$ production. 
All the above contributions would largely cancel in a ratio 
for simultaneously stored nuclei, with the partial exception of the luminosity.
In practice, for an integrated luminosity of 1~pb$^{-1}$, the statistical
uncertainty is expected to dominate.

\begin{figure}[ht]
\vspace{-1.3cm}
\begin{center}
\leavevmode
\hbox{%
\hspace*{-0.8cm}
\epsfxsize = 10cm
\epsffile{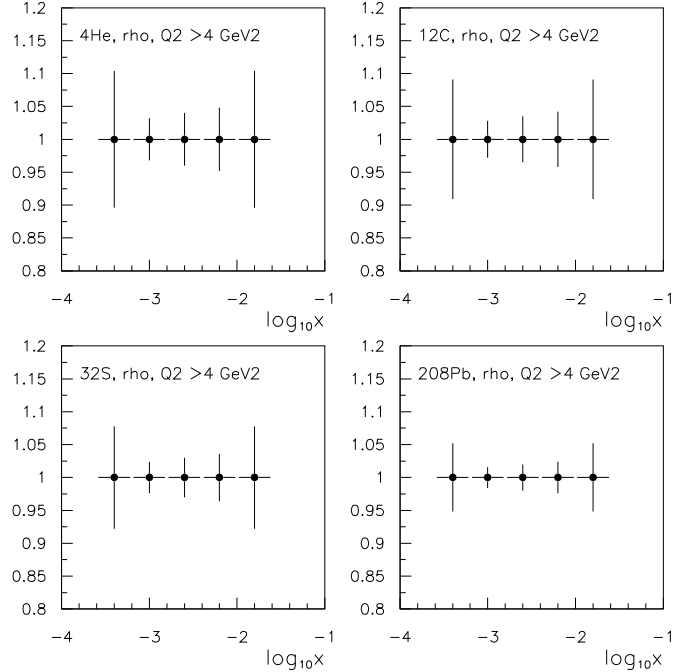}}
\end{center}
\vspace{-1cm}
\caption{Expected statistical accuracy (vertical error bars) on 
the ratio ${R_V^{1/2}}$ for 
elastic production ($Q^2>4$~GeV$^2$) of $\rho$ mesons.
In all cases a luminosity of 1~pb$^{-1}/A$ for each nucleus was 
assumed. The horizontal bars indicate the size of the bins.}
\label{rho_el}
\end{figure}

Fig.~\ref{rho_el} shows the expected statistical uncertainty
for $\rho^0$ production at $Q^2>4$~GeV$^2$.
Here as well we used DIPSI, and the parameters were chosen 
so that the generator describes the ZEUS data~\cite{rhodis93}. 
Events were generated in the range $4<Q^2<100$~GeV$^2$ and
$30<W<300$~GeV. The $\rho^0$ was assumed to decay into 
$\pi^+\pi^-$ pairs; pairs with masses between 0.3 and 1.4~GeV were
considered. Events in which the decay pions 
were outside the coverage of the tracking detectors
of H1 and ZEUS (approximately $15^{\circ} < \vartheta < 164^{\circ}$)
were rejected. A further reduction in the number of accepted events 
by a factor 0.6 was applied to account for efficiency and acceptance 
effects; this factor was independent of $\bar x$.

One can see from the figures that the accuracy of the measurements would be 
sufficient to discriminate between the colour transparency expectation
of a ratio close to unity and the vector meson dominance expectation of
the ratio $R{_V}\approx A{^{-2/3}}$;  cf. also the discussion in 
section~\ref{Colour transparency phenomena}.

\subsection{Parton Propagation in Nuclei} 

This has been studied in various fixed target experiments in the past
\cite{EMChads,E665h}.  It 
is investigated by measuring the final state hadrons in the deep inelastic 
scattering.  Since there will be many more than one hadron per event in 
experiments at HERA 
the statistical errors are not expected to be a limitation 
and high accuracy should be possible.  
This subject is considered in more detail in section~\ref{partonprop} and 
in \cite{Pavel}.

\subsection{Other Physics}

In addition to the primary programme for nuclear beams in HERA outlined above 
there will be additional experiments which have not been explored in these 
studies.  For example, 
there is physics interest in studying photoproduction from nuclear 
targets in the HERA energy range as well as sensitivity to physics beyond 
the standard model.  Such sensitivity arises in $e$-nucleus collisions via 
coherent two photon interactions allowing production of positive C parity 
states which will be inhibited at LEP.  This is discussed by Krawczyk and 
Levtchenko  \cite{KandL}.  

\section{Theoretical Overview}

\subsection{Introduction}

In this overview we describe those aspects of the phenomenology of QCD 
which lead to the nonlinear effects referred to earlier.   These effects 
are thought to be related to the mechanisms which will eventually 
limit the growth of the nucleon structure function as $x$ decreases 
at finite $Q^2$.  

\subsection{Space-time Picture of DIS off Nuclei at Small $x$}
\subsubsection{The Rest Frame}
 In the rest frame of
the target nucleus
 the life-time of a fluctuation
is given by the formula
\begin{equation}
\tau = \frac{\beta}{m_N x_{Bj}},      \label{1}
\end{equation}
where  $\beta=Q^2/(Q^2+M^2) < 1$. $M$ is the mass
of
the $q\bar q$ system
\begin{equation}
M^2=\frac{k_t^2+m_q^2}{z(1-z)},    \label{2}
\end{equation}
where $z$ is the light-cone momentum fraction, $k_t$  the transverse 
momentum  and
$m_q$  the mass of the quark.
 Perturbative QCD studies  show that
the most probable configurations are those for which $M^2 \approx Q^2$.
In the case of transversely polarised photons both configurations with small
 $k_t$ and highly asymmetric fractions $z$ and configurations with comparable
$z$ and $1-z$ contribute to the cross section. For the case of the 
 longitudinal photons the asymmetric contribution is strongly suppressed.

In the language of noncovariant diagrams this corresponds to
 the virtual photon
fluctuating  into a quark-antiquark pair at a
 longitudinal distance $l_c={\beta\over  m_Nx}$
from the nucleus which far exceeds the nuclear radius. 
The distance $l_c$ is referred to as the ``coherence length''.
 The pair 
 propagates essentially without transverse expansion until it reaches
 the target. 
QCD evolution leads to a logarithmic 
decrease of $\beta$  with increasing $Q^2$.
At HERA coherence lengths of up to 1000 fm are possible, so that the
 interaction of the $q\bar q$ pair with nuclear matter can be
 studied in detail -- notably its transparency to small size pairs -
 {\it colour transparency}.

\begin{figure}
\begin{center}
\vspace{-2.4cm}
\leavevmode
\epsfig{file=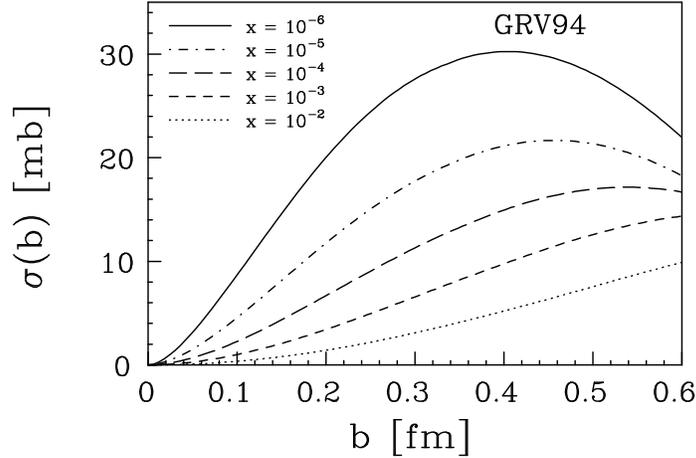,height=11cm,bbllx=19 ,bblly=142 , bburx=577 ,bbury=700}
\end{center}
\vspace{-2.1cm}
\caption{Colour-dipole cross section, $\sigma_{q \bar q N}(x,b)$ of Eq. 
(\protect\ref{sigmab}), as a function of the transverse size of the 
$q\bar q$ pair for various values of $x$ and for the GRV94 parametrization
 of the nucleon's gluon density.}
\label{sigmabb}
\end{figure}

 At HERA new features of colour transparency
should emerge: the 
incident  small size $q\bar q$ pair resolves  small $x$ gluon fields 
with virtualities $\sim Q^2$. If the transverse size of the $q\bar q$ pair
is  $r_t=b_q-b_{\bar q}$ , the 
 cross section for interaction with a nucleon is
\cite{BBFS93}
\begin{equation}
\sigma_{q\bar q,N}(E_{inc})={\pi^2\over 3}r_t^2\alpha_s(Q^2)xg_N(x,Q^2),
\label{sigmab}
\end{equation}
where $Q^2 \approx {\lambda(x)\over r_t^2}, 
\lambda(x \approx 10^{-3})\approx  9,
x={Q^2\over 2m_NE_{inc}}$.  Since the gluon density increases
 rapidly with decreasing $x$, even small size  pairs may 
interact strongly, leading to some sort of {\it perturbative colour opacity}
-- the  interaction of a small object with a 
large object with a cross section comparable to the geometric size of the 
larger object  (Fig.~\ref{sigmabb}). 

Unitarity considerations for the scattering of a small size  system \cite{FKS}
 -- i.e. the  requirement that 
$\sigma_{inel}(q\bar q, target) \le \pi R^2_{target}$ --
indicate that nonlinear effects
 (i.e. effects not accounted for by the standard evolution
 equations) should become significant at much larger 
values of $x$ in $eA$ scattering than in $ep$ scattering.

\subsubsection{The Breit Frame}

\begin{figure}
\begin{center}
\leavevmode
\epsfig{file=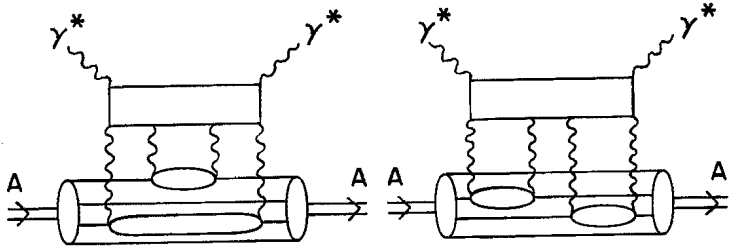,height=5cm}
\vspace{-1.3cm}
\end{center}
\caption{Typical fan diagrams leading to nonlinear evolution of $g_A(x,Q^2)$.  
}
\label{mq2}
\end{figure}

In the Breit frame, small $x$ partons 
in a nucleon are localized within a longitudinal distance
$\sim 1/xp_N$, while the distance between two nucleons is
$\sim r_{NN}m_N/p_N $ ($r_{NN}$ is the distance between nucleons in the
rest frame and $p_N$ is the nucleon momentum). 
Therefore partons 
with 
$x<1/(2 m_N r_A)$, where $r_A \approx r_0 A^{1/3}$ fm  is the nuclear 
radius and  $r_0=1.1$ fm cannot be localized to better 
than the whole nuclear longitudinal 
size. Hence low $x$ partons emitted by  different 
nucleons in a nucleus can overlap spatially and fuse,  provided the 
density is high enough, leading to 
shadowing of the 
partonic distributions in bound nucleons with respect to the free 
nucleon ones and to nonlinear effects already at values 
of $x \sim 10^{-4} \div 10^{-3}$. 
For example, in the simplest model of nonlinear effects corresponding to the
fan diagrams of Fig.~\ref{mq2}, the additional contribution $\delta g_A(x,Q^2)$  to 
$g_A(x,Q^2)$ due to the nonlinear term in the equation
for the $Q^2, x$  evolution of the gluon density is~\cite{MQ}:
\begin{equation}
Q^2 {\partial \over \partial Q^2}{ \delta x g_A(x,Q^2) \over A}=
-{81 \over 16}{A^{1/3}\over Q^2r_0^2}\alpha_s^2(Q^2)\int^1_x{du\over u}\left[
ug_N(u,Q^2)\right]^2.
\label{MQ}
\end{equation} 
The analogous equation for the gluon density in the nucleon has a much smaller
 coefficient -- approximately  by a factor $r_0^2/r_N^2 A^{1/3}$,
 where $r_N \sim 0.8$ fm is the nucleon radius. Once again one can see 
then that
the $x$-range where nonlinear effects
 become significant differs   for a heavy nucleus and for a nucleon  
by more than two orders of magnitude, assuming 
 $xg_N(x,Q^2) \propto x^n$ with $n\sim -0.2$.  

Thus electron-nucleus collisions at HERA
can  be seen as efficient amplifiers of nonlinear QCD effects.

\subsection{Theoretical Framework for Small $ x$  Phenomena in $eA$ 
Collisions}
\subsubsection{Perturbative and Nonperturbative Shadowing}
At small $x$ the DIS cross section per nucleon in a nucleus is smaller for a
 bound nucleon than for a free one, the so called shadowing phenomenon. 
Shadowing is  determined by a combination of 
non-perturbative and perturbative effects. In the DGLAP evolution
equations  one can 
express shadowing at large $Q^2$ through the shadowing at the normalization
point $Q_0^2$. This type of shadowing is connected to 
the soft physics.
It can be visualized e.g. in the aligned jet 
model of Bjorken~\cite{BJ71}, extended to account for QCD evolution effects
\cite{FS88}.
  A virtual photon converts to a $q \bar q $ pair with
 small transverse momenta (large transverse size) which interacts with 
the nucleus with a 
hadronic cross section, leading to shadowing.
  The effective small phase volume 
of these configurations ($\propto {\lambda \over Q^2}$)
leads to Bjorken scaling
and it is due to colour transparency \cite{FS88}.

 At large $Q^2$, these $q \bar q$ pairs
  evolve into systems with gluons, leading to a 
shift of shadowing to smaller $x$, which is equivalent to the standard
 $Q^2$ evolution of parton distributions. These $q \bar q$ pairs,
 which interact with the target nonperturbatively,
 seem to be responsible for most of the shadowing at 
intermediate $Q^2$ and $x \sim 10^{-2}$ which has been 
studied at fixed target energies.
This mechanism of shadowing is effective for $\sigma_T$ only since for 
$\sigma_L$ the aligned jet contribution is strongly suppressed. 
For 
$\sigma_L$ (as well as for the production of heavy quarks) one is
 more sensitive to the shadowing due to the  interaction
 of small size $q \bar q$
 pairs with  the nuclear gluon field which can be shadowed.
  
\begin{figure}
\begin{center}
\leavevmode
\epsfig{file=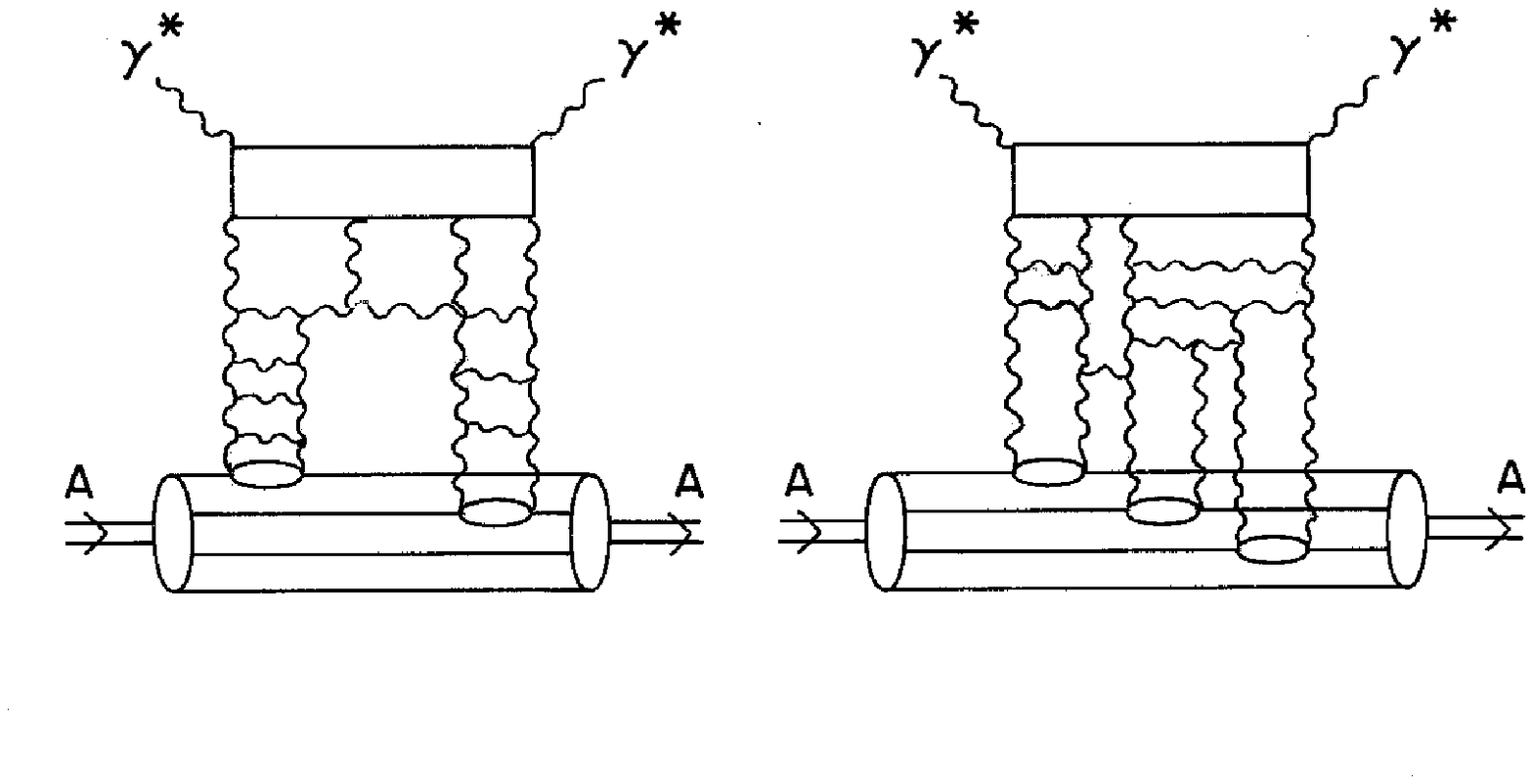,height=5cm}
\vspace{-1.3cm}
\end{center}
\caption{Examples of typical perturbative QCD diagrams contributing to 
 nuclear shadowing.}
\label{horder}
\end{figure}

At smaller $x$ the situation may change rather dramatically because,
as the recent HERA data indicate, already for $Q^2
\sim $1.5 GeV$^2$ at $x \sim 10^{-4}$ perturbative contributions to
$F_{2p}(x,Q^2)$ appear to become important,
leading to a rapid increase of the structure functions with decreasing  
$x$.  Hence  contributions of various  
perturbative mechanisms which may generate shadowing for configurations of
 a size smaller than the
 hadronic size  may become important. Perturbative QCD may
 be applicable to those small size pairs.  Typical contributions 
involve  diagrams of the eikonal type, various enhanced diagrams, etc. 
 (Fig.~\ref{mq2},\ref{horder}). 

\subsubsection{Shadowing and Diffraction}
\label{Shadowing and diffraction}
In practically all models
 it is  assumed that nuclei are built of nucleons. So the condition that the 
matrix element $<A|T[J_{\mu}(y)J_{\nu}(0)]|A>$ involves only nucleonic
initial and final states is implemented\footnote{ The condition
that nuclei are built of nucleons is not so obvious in the fast frame picture.
However it is implemented in most of the models \cite{MQ,McLerran}.}.
Under these natural assumptions one is  essentially not sensitive to 
any details of the nuclear structure, such as short-range correlations etc.

 In the case of scattering off the deuteron
and light nuclei the same diagrams contribute to the cross section for 
 diffraction in  $ep$ scattering and the  cross section  for shadowing --
 hence similar nonlinear phenomena like those described by eq.(\ref{MQ})
 are involved  in each case.  
For example for the deuteron~\cite{Gribov69}:

\begin{equation}
\sigma_{shad}={\sigma_{tot}(eD)- 2\sigma_{tot}(eN)\over \sigma(eN)}=
{ {d\sigma_{diff}(ep)\over dt}
_{\left|t=0\right.}\over
\sigma_{tot}(ep)}{1 \over 8 \pi R_D^2}R,
\end{equation}
where $R={(1-\lambda^2)\over (1+\lambda^2)}$, 
$\lambda =Re A/Im A \approx {\pi \over 2}{\partial \ln A \over \partial \ln s}
$ for the amplitude
$A$ of $\gamma^*p$ scattering and $R_D$ is the deuteron radius.  
 For small $x$, $\lambda$ may be as large as 0.5, leading
to $R \sim 0.5$ especially for the case of the longitudinal
cross section. So already for 
light nuclei the study of the total cross sections
of scattering from nuclei would allow 
{\it to establish  a fundamental connection between
the two seemingly unrelated phenomena of diffraction at small $t$ in $ep$
scattering
and nuclear shadowing}.
With the increase of $A$ more complicated nonlinear interactions with several 
nucleons become important, see  e.g.  Fig.~\ref{horder}b.

Nuclear shadowing for the total cross sections has a simple physical meaning -
it corresponds to a reduction of cross section due to screening of one 
nucleon by another (as well as 
by several nucleons for $A >2$). 
If one treats the deuteron as a two nucleon system it is 
possible to apply the Abramovskii, Gribov,  Kancheli (AGK)
cutting rules \cite{AGK} to elucidate {\bf the connection between  
  nuclear shadowing,  
diffraction and fluctuations of multiplicity.} One observes that the 
simultaneous interaction of the $\gamma^*$ with the two nucleons of the 
deuteron modifies not only 
the total cross section but also the composition of the produced  final 
states.  It increases the cross section for diffractive scattering off 
the deuteron due to diffractive scattering off both nucleons by 
$\delta \sigma_{diff}=\sigma_{shad}$.  At the same time the probability 
to interact inelastically with one nucleon only is reduced since the 
second nucleon screens the first one: $\delta \sigma_{sin
gle}=-4 \sigma_{shad}$. In addition, a new process emerges in the case of 
the deuteron which was absent in the case of the free nucleon - 
{\it simultaneous} inelastic interaction with both nucleons 
which leads to a factor of two larger multiplicity densities
 for rapidities away from the current fragmentation 
region: $ \sigma_{double}=2 \sigma_{shad}$.  
Altogether these contributions constitute
$-\sigma_{shad}$,  the amount by which the total cross section is reduced
\footnote{For simplicity we 
give  here relations for the case of purely imaginary $\gamma^* N$ 
amplitude
 ${Re A \over Im A} =0$. }.

To summarize,  {\it there is 
a deep connection between the phenomena of diffraction observed at HERA 
in $ep$ scattering  and nuclear shadowing  as well as the $A$-dependence 
of diffraction  and the distribution of the multiplicities in DIS}. 

It follows from     
the above discussion that it is possible to get information 
about the dynamics of nuclear shadowing and hence about nonlinear effects 
by studying several {\bf key DIS phenomena} such as:  nuclear shadowing 
for inclusive cross sections $F_2^A, {\sigma_L \over \sigma_T}, 
F_2^{A charm}$;  the cross section for nuclear diffraction;  
the multiplicity distribution for particle production 
in the central rapidity  range; diffractive  
 production of vector mesons.  The advantage of the latter process is that 
one gets a rather direct access to the interaction of a 
small colour dipole with matter. It is in a sense an exclusive analogue of 
$\sigma_L$ which is easier to measure.

\subsection{The $A$-Dependence of Parton Distributions at Small $x$ }

As discussed above, the nucleus serves as an amplifier for  
nonlinear phenomena expected in QCD at small $x$.
 The simplest example of such effects is given by equation
 (\ref{MQ}) where the nonlinear term is 
proportional to the square of the nucleon gluon density.
If  shadowing were absent the parton densities per unit transverse area
would be  enhanced by a factor
$A^{1/3}$ as compared to the free nucleon case. Hence even just an upper limit 
on the parton densities based on unitarity -- that the cross section for the 
inelastic interaction of a small dipole with a nucleus may not exceed 
$\sigma_{inel}=\pi R_A^2$ --
leads to the  expectation of
nonlinear phenomena -- shadowing of an observable magnitude -- already at
$x \sim 10^{-3} \div 10^{-4}$ \cite{FKS}. 

Hence, from detailed studies of the $A$-dependence of the parton densities
 it would be
possible both to check the dominance of the two-nucleon screening mechanism
for $x \sim 10^{-2}$ \cite{FS89,KP}
 and to extract information about the coherent interaction of
the virtual photon with three (four) nucleons at $x \le 10^{-3}$.

For $x\ge  10^{-2}$ for any nucleus and for all $x$ for light nuclei, the
 main contribution to shadowing is given
by the interaction with two nucleons of the target. Hence in this regime
there is a relatively simple connection with the diffraction of a virtual photon 
off a proton  -- which is the simplest nonlinear effect in the 
perturbative domain in QCD. 
For smaller $x$ and heavy nuclei,
when essential longitudinal distances become comparable and ultimately
exceed the diameter of the nucleus,
several nucleons at the same impact parameter contribute to
the screening. It is worth emphasizing that these multi-vacuum exchange
processes cannot be singled  out unambiguously
 using a nucleon target. 
The relevant QCD diagrams for the total cross section of
 $\gamma^*A$ interaction
  are rather similar to 
higher-order nonlinear diagrams 
for the proton target -- except that in the nuclear case
one has to impose the condition that
 couplings to the individual nucleons are colour singlets, see e.g.
Fig.~\ref{mq2},\ref{horder}. 

In a sense, the studies of nuclear shadowing at small $x$ and large $Q^2$
can be considered as a simpler model of 
nonlinear  effects which occur in the case of a nucleon target. 
In the latter case it is not easy to relate the 
coupling of say two vacuum exchanges (or 
a ladder with 4 gluons in the $t$-channel) with a nucleon to the coupling 
of one vacuum exchange with a nucleon.
In fact the region of $10^{-3}\ge x \ge 10^{-4}$ may be optimal in this respect
since nonlinearities for the nucleon case are still small though
nonlinearities for the nuclear case are  already quite substantial.
It is worth emphasizing that experience of the studies of the total 
hadron-nucleus scattering indicates that interaction with bound 
nucleons for the total cross sections can well be approximated 
by the interactions with free nucleons (for a recent analysis see 
\cite{Jenmil}). Therefore {\it nuclear structure effects do not obscure 
the interpretation of nuclear shadowing effects}. 

\begin{figure}
\vspace{-0.3cm}
\centerline{
\epsfig{file=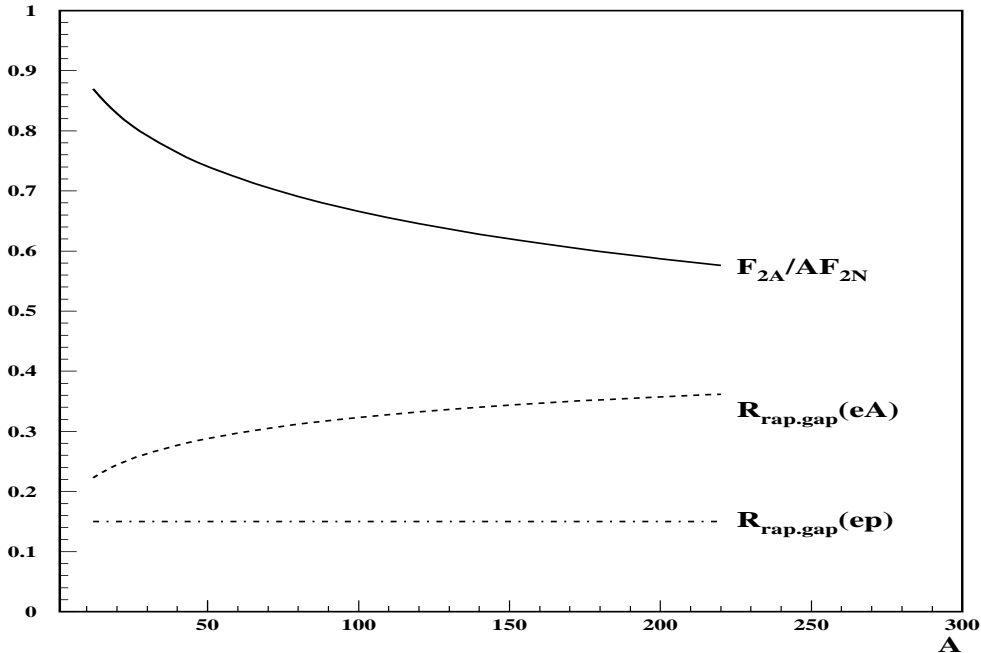,width=15.0cm,height=10.0cm}}
\vspace{-0.3cm}
\caption{$A$-dependence of nuclear shadowing  
and probability of rapidity gap events in the colour screening model
of shadowing; dot-dashed curve assumes $A$-independent probability
of rapidity gap events.}
\label{shaddif}
\end{figure}

Using current information from HERA on diffractive production in $ep$ 
 scattering it is straightforward to  estimate  the amount of nuclear 
shadowing at small $x$  taking into account interactions with 3 or more  
nucleons using the
eikonal approximation with an effective cross section determined from 
diffractive data, see eq.(\ref{sigmaeff}) below.
The result of the calculation \cite{FSAGK}
is shown in Fig.~\ref{shaddif} for $Re/Im=0$; for $A \ge 12$  
it weakly depends on the value of $Re/Im$. Since the data on diffraction
 indicate that the fraction of diffractive events in DIS weakly 
depends on $x,Q^2$ these considerations 
show that significant shadowing effects should be present for 
$F_{2}^A(x,Q^2)$ in the wide small $x$ range of HERA. 
Note that the   shadowing effect in DIS is expected to be much
smaller than for the case of real photon scattering since the effective 
cross section for interaction of the hadron component of quasi-real photon 
at HERA is a factor of $\sim 3$ larger than for a highly virtual photon 
(we use here the HERA data on diffraction for  real photons \cite{H1}).  

Since the interaction of the octet colour dipole $gg$ is a factor of $9/4$ 
stronger than for the $q \bar q$ dipole, nonlinear effects are expected 
to be more important for  gluons.  
So gluon shadowing would provide even more direct access to 
nonlinear phenomena. Note that in this case there is no simple
 relation of shadowing with diffraction in $\gamma^* +p$ DIS, so 
any information about gluon shadowing would be complementary to
 the information from $ep$ DIS. 
There are very few data on the 
gluon distribution in nuclei. Among them, the enhancement of the 
gluon distribution
at $x \sim 0.1$ indicated by the inelastic $J/\psi$ 
production data~\cite{nmc2}. Also
the  analysis \cite{GP} of the scaling violation for  the ratio 
$F_{2}^{Sn}/F^C_{2}$~\cite{nmc_q2dep} under the assumption that higher 
twist effects are not important in the $Q^2,x$ range of the data allows to 
extract information about the $A$-dependence of gluon distributions, 
indicating some nuclear shadowing for 
$G_A$ for $x \le 0.01$ and an enhancement at $x \sim 0.1$,
 see Figure in \cite{GP}\footnote{The shadowing for gluons should be 
accompanied by a significant enhancement at larger $x$ since
 the total momentum fraction carried by gluons in nuclei is not
 suppressed and is probably slightly enhanced~\cite{FLS}.  }.  
Theoretical expectations for gluon shadowing discussed in the 
literature are quite different -- from a larger effect than for 
$F_{2}^A$~\cite{FLS93}, to an effect comparable to that of 
quarks \cite{Qiu,FLS,Escola,EQW}
 to substantially smaller shadowing \cite{NZ91};
 see also contributed papers
to these proceedings.

Comparison of different determinations of 
shadowing of gluons  and measurements of the scaling violation 
for the $F_2^A/F_2^D$ ratios will allow to determine 
the range of applicability of  the DGLAP evolution equations and hence
 provide unique clues to the role of nonlinear effects.

It is worth emphasizing also that knowledge of parton 
distributions in bound nucleons at these values of $x$
will be crucial also for studies in heavy-ion 
physics at the LHC and RHIC.

\subsubsection{BFKL Pomeron} 
One can envision several strategies 
 for the study of the BFKL Pomeron in DIS. The main requirement is to enhance 
the contribution of scattering of small transverse size configurations 
in the ladder. It is natural to expect that screening for these configurations 
would be minimal. Hence for heavy nuclei the contribution of the BFKL 
Pomeron can be enhanced.  

\begin{enumerate}
\item 
A procedure can be envisioned to 
study the $A$-dependence of $F_{2}^A(x,Q^2)$ at  small $x$ to extract the term in
 the structure functions $\propto A$ and then study its $x$ dependence. 
  Based on the above argument A.Mueller has suggested \cite{Mueller} 
  that 
the $x$ dependence of this term (linear in $A$) would be  closer to BFKL 
type behaviour.
 \item 
One promising direction to look for the BFKL pomeron is the Mueller-Navalet 
process of producing two high $p_t$ jets with large a rapidity difference
\cite{MN}  to 
suppress the contribution from
small transverse momenta (large transverse distances)
 in the ladder.
 In the case of a nuclear 
target  large distance contributions would be screened out to a large 
extent.

\item 
Another   possibility is the 
production of $\rho$ mesons at large $|t|$
in inelastic diffraction. To enhance the contribution of the 
BFKL Pomeron it is desirable to increase the contribution of 
small configurations in the $\rho$ meson, i.e. quark-antiquark pairs
with small transverse separation. Large size configurations 
can be filtered out by exploiting the fact that they are absorbed on the 
nucleus surface. Once again extracting the term in the cross section 
$\propto A$ would allow to enhance the contribution of the BFKL Pomeron.

\end{enumerate}

\subsection{Diffraction off Nuclei}
\subsubsection{Introduction}

Diffraction studies have been defined as one of the primary goals of 
nuclear beams in HERA.  Such processes 
can be interpreted using two complementary languages
depending on whether the rest frame of the nucleon
 or the Breit frame are used:

$\bullet$
 Scattering of electrons on colourless components
of the proton \cite{H1F3D,ZEUSF3D}. Such scattering
 may be  identified, for the very  low $x$ events dominated by diffraction,
 with the interaction with the vacuum $t$-channel 
exchange   which is often referred to as the Pomeron, 
 $I\!\!P$. This object is not necessarily the same as the
 Pomeron of the Gribov-Regge high-energy soft interactions 
(see report of the diffractive group).
Deep inelastic electron scattering leading to the presence
of a rapidity gap can thus be considered as probing  the internal parton
structure of the  $I\!\!P$ originating from the proton.

One of the questions of primordial importance which may be addressed
within the future electron-nucleus scattering program at HERA is then 
``how universal is the internal structure of the Pomeron?" or,
more precisely: ``Is the internal structure of the Pomeron originating
from various hadronic sources (protons, neutrons, nuclei) the same?".
We shall show below how nuclei may help in answering these questions.

$\bullet$ The diffractive interaction  of different hadronic components of the
virtual photon with the target via vacuum exchange. Diffraction
 predominantly selects the $\gamma^*$ components which interact with
 sufficiently large cross sections such as large transverse size $q \bar q$,
$q \bar q g$ colour dipoles.
Therefore the study of diffraction plays a very important role
 in determining the relative importance of small and large size configurations
and addressing the question
whether
small white objects interact weakly or not. Indeed if the interaction with
a target becomes sufficiently strong at small impact parameters
the  cross section for  diffraction
 (which includes both elastic  scattering and 
inelastic diffractive dissociation)
would reach the black body limit of 50\% of the total cross section.

\subsubsection{Theoretical Expectations}

Diffraction off a nucleon (including dissociation of the nucleon)
constitutes about $15$-$20\%$ of the deep inelastic events.
Therefore  the interaction is definitely far from being
 close to the scattering 
off a black body. Even this 
number came a surprise in view of the large $Q^2$ value
 involved. Using the  generalized 
optical theorem as formulated by Miettinen and Pumplin, 
one can estimate the effective total cross section for the 
interaction of the hadronic components
of the $\gamma^*$ as  

\begin{equation}
\sigma_{eff}=
16\pi{ {d \sigma_{diff}^{\gamma^*+p \rightarrow X +p}\over dt}_{\left|t=0\right.}
\over 
\sigma_{tot}(\gamma^*N)}
\approx 12\div 15 \mbox{mb}.
\label{sigmaeff}
\end{equation}
This cross section is significantly smaller 
than the $\rho N$ cross section which at the HERA energies can be estimated
to be close to 40 mb
using the vector dominance model and the Landshoff-Donnachie fit \cite{Lan}:
\begin{equation}
\sigma^{\rho N}_{tot}(s)=
\sigma^{\rho N}_{tot}(s_0){\left(s \over s_0\right)}^n,
\label{LD}
\end{equation}
where $n \approx 0.08, s_0=200$ GeV$^2, \sigma^{\rho N}_{tot}(s_0)=25~$ mb.
However it is sufficiently large to result in a 
substantial cross section 
 of diffraction for small 
$x$ -- it can reach 30-40\% for large $A$ (Fig.~\ref{shaddif})\cite{FSAGK}. 
For large $A$
the coherent diffraction 
dominates 
when the incoming wave is sufficiently 
absorbed at small impact parameters  which, by virtue of 
Babinet's principle, corresponds to scattering beyond the nucleus. In 
such processes the nucleus remains intact and the average momentum 
transfer is very small ($
\left<t\right>
\propto A^{-2/3}$).

One expects that hadronic configurations interacting 
with different strength contribute to diffraction (cf. Fig.~\ref{sigmabb}).
The  parameter
$\sigma_{eff}$ characterizes  just the average strength of this interaction,
while the distribution over the strengths is expected 
to be quite broad. The study of diffraction off nuclei
allows to separate contributions to diffraction of large and small size 
configurations  due to 
 {\bf the filtering phenomenon}: with the increase of $A$ the 
relative  contribution  of more weakly interacting
(smaller size) configurations  should 
increase since they are less shadowed, 
leading to a relative enhancement of the colour transparent subprocesses.

Examples of   promising processes are:
\begin{itemize}
\item Diffractive production of charm.
The $A$-dependence of this process would be interesting 
already at low $Q^2$ since the essential 
transverse distances are, naively, of the order of $1/m_c$, where $m_c$ is the
charm quark mass.
Since the cross section for the interaction of a colour dipole of such size
is small for $x\sim 10^{-2}$, the cross section for diffractive charm production
at these values of $x$ is small and practically not shadowed, 
leading to a cross section $\propto A^{4/3}$.
At the same time the $c\bar c-N$ cross section 
  increases rapidly with decreasing $x$ (increase of energy) for fixed $Q^2$.
Therefore at  HERA energies diffractive charm production  in $ep$ 
collisions may become a
substantial part of the total diffractive cross section. At this point
one expects the emergence of 
shadowing in diffractive charm production in $eA$ collisions, leading to
 slowing down of the $A$-dependence of
 diffractive charm production as compared 
with the $A$ dependence at $x \sim 10^{-2}$.

One can go one step further and study  the $A$-dependence 
of $p_t$ distributions for diffractive charm production. 
Smaller size components
will be less absorbed and so their relative contribution may increase with $A$.

\item Diffractive production of two high $p_t$ jets.

Selection of large $p_t$ jets enhances the contribution of diffraction 
of small size configurations. 
Hence, one expects broader $p_t$ distributions
in the case of nuclear targets 
(smaller jet alignment) with nontrivial dependences on $W$ and $Q^2$.
For example, if  we fix the $p_t$ of the jets,
the $A$-dependence of dijet production should become
weaker with increasing energy  reflecting the increase of the absorption 
(which can be studied this way).
If on the other hand we fix $W$ and consider the $A$-dependence as
 a function of 
$p_t$, a stronger  $A$-dependence is expected. 
Effectively, this would be another 
way to approach colour transparency via filtering out of the soft components.

\end{itemize}

To summarize, a
 study of inclusive diffraction will give better insights into 
 the structure of the Pomeron. The
interplay of soft and hard contributions will lead to a 
breakdown of factorization for the structure function of the Pomeron.
Stated  differently, the check of the degree 
of ``universality" of the Pomeron -- i.e.
whether the ``nuclear" Pomeron is different from the Pomeron observed in $ep$ 
diffraction -- will provide a very sensitive test of QCD dynamics.

An important aspect of the diffractive studies is the colour transparency
phenomenon. In view of its special interest we will discuss this separately 
below.

\subsubsection{Multiplicity Fluctuations}  

\begin{figure}
\centerline{
\epsfig{file=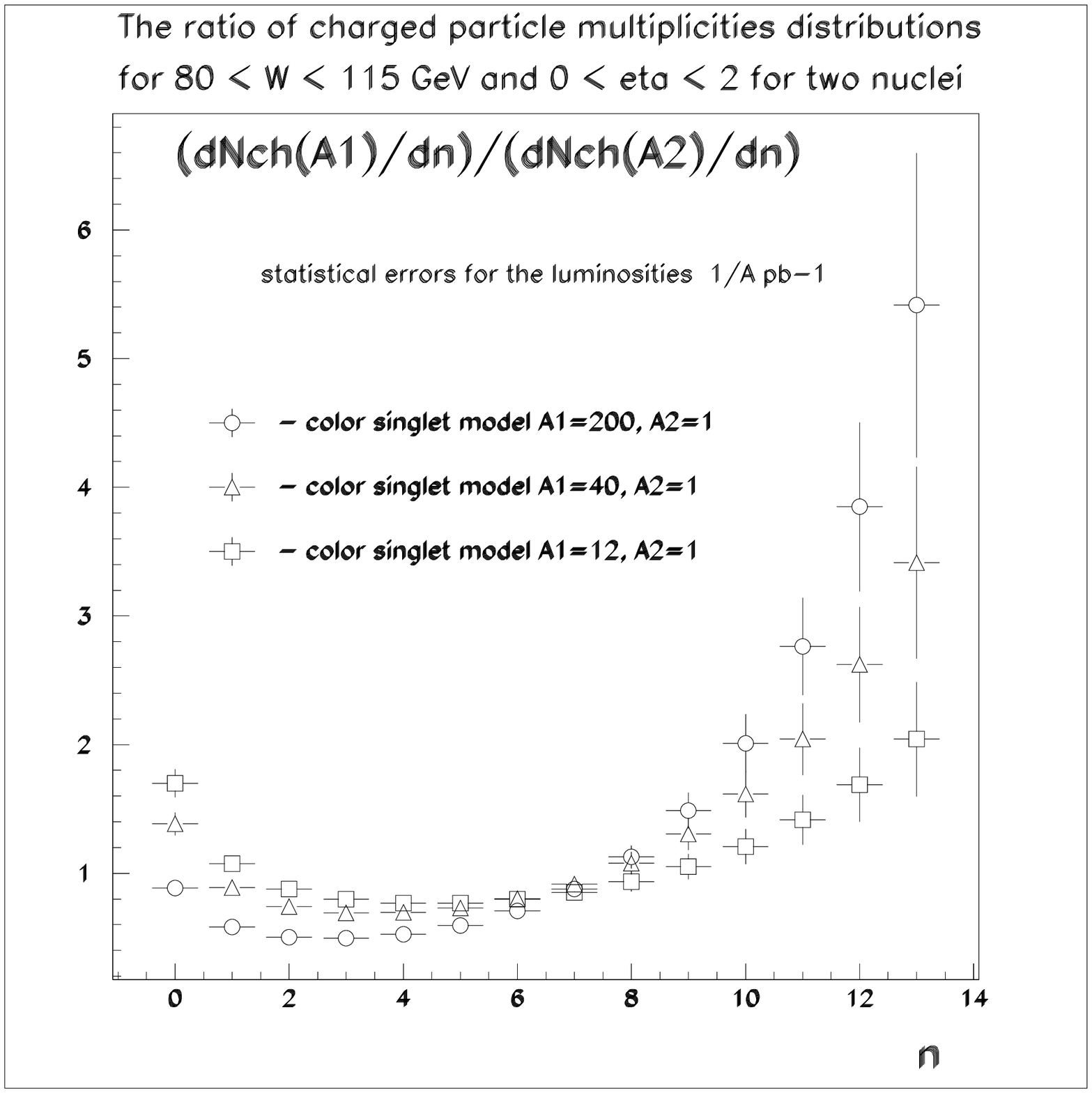,height=10.0cm,bbllx=19 ,
bblly=142 , bburx=577 ,bbury=700}}
\caption{$A$-dependence of distribution over multiplicity calculated in the 
colour singlet model for a luminosity of 1 pb$^{-1}$.}
\label{multiplicity}
\end{figure}

As we explained in section~\ref{Shadowing and diffraction} diffraction 
 off nuclei and nuclear shadowing  are
closely related to the simultaneous inelastic interactions of the virtual
photon with several nucleons. Such interactions produce events with large 
multiplicity densities in the central rapidity range, leading to 
a much broader distribution over multiplicities in  
$eA$ collisions than in the $ep$ case. Study of this effect will provide 
information complementary to that obtained from nuclear shadowing 
about the structure of the vacuum exchange at small $x$.
Interesting phenomena to look for may be:
\begin{enumerate}
\item Local fluctuations of multiplicity in the central rapidity region, 
e.g. the observation of a broader distribution of the number of particles per 
unit rapidity, $n(\Delta \eta)$, than in $ep$ scattering \cite{FSAGK},
 see Fig.~\ref{multiplicity}.  

\item Long range rapidity fluctuations 
-- i.e. positive correlation of the increase of multiplicity 
in one rapidity interval  with the  increase
 of multiplicity several units away.

\item Correlation of the central multiplicity with the multiplicity of 
neutrons in the 
neutron detector (most effective for heavy nuclei).
\end{enumerate}

\subsection{Colour Transparency Phenomena}
\label{Colour transparency phenomena}

An important property of QCD is that small objects are expected
 to interact with hadrons with small cross section \cite{Low}.
This implies that in the processes dominated by the scattering/production of 
hadrons in ``point-like''(small size) configurations (PLC)  when
 embedded in the nuclei, the  projectile or the outgoing hadron 
essentially does not 
interact with the 
nuclear environment \cite{Bro82,Mue82}.
In the limit of colour transparency one expects for an incoherent cross
 section  a linear dependence on $A$, for example
\begin{equation}
{d \sigma (e+ A\rightarrow e + p +(A-1)^*) \over d Q^2}=Z 
{d \sigma (e+ p\rightarrow e + p)\over d Q^2},
\end{equation}
while for coherent processes at $t=0$ one expects
\begin{equation} 
{d \sigma (\gamma^*+ A\rightarrow X +A) \over d t}=A^2 
{d \sigma (\gamma^*+ N\rightarrow X +N) \over d t}.
\end{equation}
No decisive experimental
tests of this property of QCD  were performed so far since in most of the 
 current experiments the energies were not sufficiently high
 to prevent  expansion of the produced small system.
The high-energy E665 experiment \cite{E665rho}
 at FNAL has found some evidence for
colour transparency in the $\rho$ meson production off nuclei. 
However, the data have low statistics, 
cover a small $x,Q^2$ range and cannot reliably separate events without 
hadron production.

\begin{figure}
\begin{center}
\leavevmode
\epsfig{file=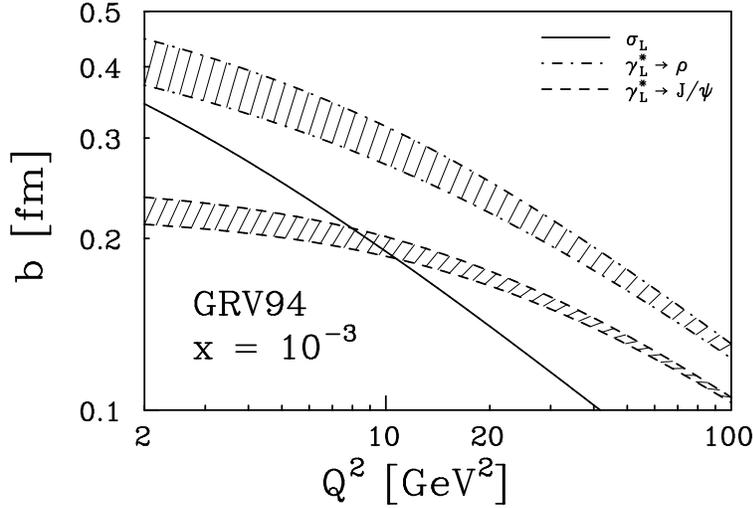,height=7cm}
\end{center}
\vspace{-0.7cm}
\caption{Average transverse size of the $q\bar q$ components effective in
${\cal A}_{\gamma^*_LN \rightarrow VN}$  for $\rho-$ and $J/\psi$-meson production and 
$\sigma_L$. The probed $Q^2$ scale is inversely proportional to $b^2$.}
\label{bq2}
\end{figure}

A quantitative formulation of colour transparency for high-energy processes
can be  based on eq.(\ref{sigmab}). For the case of nuclear targets it
 implies that  for a small enough colour dipole, the cross section of its
 interaction with nuclei
 is proportional to 
$A$ up to the gluon shadowing factor. As a result the 
colour transparency prediction  for 2 jet and vector meson diffractive
production   is \cite{FMS93,Brod94}
\footnote{In writing eq.(\ref{ratios}) we neglect the 
difference of $Q^2$ scales for different processes which is reflected in a 
 different dependence of the essential transverse size of the 
$q \bar q$ state on the process (see Fig.~\ref{bq2}). For a discussion of the 
 appropriate scale for dijet production see \cite{Bartels2jets}.  }:  
\begin{equation}
{{d\sigma\over dt}(\gamma^* A \to 2 jets + A)\big\vert_{t=0}\over
{d\sigma\over dt}(\gamma^* N \to 2 jets + N)\big\vert_{t=0}} =
{{d\sigma\over dt}(\gamma^* A \to V A)\big\vert_{t=0}\over
{d\sigma\over dt}(\gamma^* N \to V N)\big\vert_{t=0}} =
\left [{F^L_A(x,Q^2) \over F^L_N(x,Q^2)}\right ]^2
= {G^2_A(x,Q^2) \over G^2_N(x,^2Q)}\ .
\label{ratios}
\end{equation}
Gluon shadowing constitutes a rather small effect for
 $x \sim 10^{-2}$ (see earlier discussion). For smaller $x$ it increases 
but it is in any case much smaller than the screening effect expected in
 the case of lack of colour transparency if the produced system interacts with
cross section comparable to $\sigma_{\rho N} \sim 30$-40 mb.
For such values
 of $\sigma$ one expects the cross section to behave as $\propto A^{4/3}$
for $t=0$ which would be possible to test using diffractive production by 
 quasi-real photons. 

{\it Coherent diffractive $\rho$,$J/\psi$-meson production}

The most straightforward test of colour transparency can be made using 
 coherent production of $\rho$ or $J/\psi$-mesons at small $t$ using nuclei 
with $A \ge 12$. The $p_t$ resolution of the current detectors is good 
enough to single out the diffractive peak which is concentrated at 
$p_t \le 0.1$ GeV.
In the higher $x$ end of the range which could be studied
at HERA for  vector meson production, $x \sim 10^{-2}$, one 
expects at large $Q^2$  nearly complete colour transparency 
since  gluon shadowing effects are rather small and  decrease rapidly
 with increase of $Q^2$, while the transverse separation,$b$, between $q$
 and $\bar q$ is of the order of 0.4 fm for $Q^2 \sim 10$ GeV$^2$
and further decreases with increase of $Q^2$ (Fig.~\ref{bq2}
\cite{FKS}
).
Study of coherent $J/\psi$ meson production would allow to probe colour transparency for
propagation of even smaller 
dipoles since $\left<b_{c\bar c}(Q^2=0)\right> \sim 0.2$ fm.

 On the other hand  as discussed earlier  at the  smallest 
values of $x$ of the HERA range,  
screening effects should start to play a role even at large $Q^2$
so a gradual disappearance
 of  colour transparency is expected -- the emergence of colour opacity.
 Noticeable screening is expected already
on the basis of unitarity constraints. 
Qualitatively one may expect that the rise of the cross
 section for vector meson production with increasing energy at fixed $Q^2$
will slow down at significantly lower energies than for the case of
 the $\gamma^* +p$ reaction.
Currently theoretical calculations of  vector meson
 production by transversely polarised photons are difficult because the 
nonperturbative large distance contribution is not as strongly 
suppressed in this case as in the longitudinal case.
If 
contribution of pairs with large transverse size 
 is indeed important for
$\sigma_T$, it would be filtered out with increasing $A$ leading to larger
values of 
$\sigma_L/\sigma_T$  for large $A$.

Let us enumerate several other effects of Colour Transparency (CT)
 in diffractive production.

\begin{enumerate}
\item Production of excited vector meson states $\rho', \phi'$.  

In the CT limit, QCD predicts a universal $A$-dependence of
 the yields of  the lowest mass  and excited  states 
(this includes the effect of gluon shadowing in eq.(\ref{ratios})). This is
 highly nontrivial since the  sizes of the excited states are 
much larger, so  one  might  expect larger absorption.
On the other hand, for lower $Q^2$ average transverse distances, $b$, 
are not small. At these distances the wave functions of ground and 
excited states differ.
So in this
$Q^2$  range the relative yields of
various mesons may depend on $A$. 

\item Production of high $p_t$ dijets.
 
The uncertainty relation indicates that 
coherent production of dijets with large $p_t$, carrying all the momentum of
the diffractively produced system 
is dominated by distances $r_t \propto {1 \over p_t}$.
Hence filtering out of soft jets is expected, leading to a broadening
of the $p_t$ and thrust distributions. At the same time the study of the
$A$-dependence of low $p_t$ jets would allow to address the question
of colour opacity.
Another feature to look for would be the distribution over the 
electron-two jet plane angle as suggested in~\cite{Bartels}. 

\item Coherent diffractive production 
  at  $-t \ge 0.1$~GeV$^2$ for $A=2,4$. 

An important question here 
is the possibility to observe the ``disappearance" of colour transparency 
for $\rho$-meson production and the  emergence of ``colour opacity''
-- due to nonlinear screening effects at 
$x \sim 10^{-4}$. 
Manifestations of CT would be the increase of the differential 
cross section  ${d \sigma \over dt}$ 
below the diffractive minimum ($|t_{min}(^4He)| \approx 0.2$~GeV$^2$
and suppression of the cross section in
the region of the 
secondary maximum. A gradual  disappearance of CT in this region
with increasing energy would appear as a very fast increase  
with energy of the secondary maximum of the $t$ distribution. Remarkably, in 
this region the cross section for the process is proportional to
$[G_N(x,Q^2)]^4$, where $G_N$ is the gluon density in the nucleon
 \cite{AFS,Koepf}. The present beam optics would allow measurements
of quasielastic processes with $^4$He in the region of the secondary maximum
($|t(^4$He$)| \approx 0.4$ GeV$^2$).  For a luminosity of
10~pb$^{-1}$ it would be possible to measure the $\rho$-meson production
 cross section up to $Q^2 \sim 10$ GeV$^2$.

\item $A$-dependence of rapidity gaps between jets  in photoproduction.  

Recently photoproduction events which have two or more jets 
have been observed in the range $135
 <W_{\gamma p}< ~280$ GeV  with the ZEUS detector at HERA~\cite{ZEUS}.
A  fraction of the events has little hadronic activity between 
the jets. The fraction of these events, $f(\Delta \eta)$, reaches
 a constant value of about 0.1 for
large pseudorapidity intervals $\Delta \eta \ge 3$.The observed
 number of events with a gap is larger than that expected on the
 basis of multiplicity fluctuations assuming the exchange of a colour singlet.
This value is rather close to estimates in perturbative QCD 
\cite{Bjorken,MT,DT} neglecting
absorptive effects due to interactions of spectator partons.
It is much larger than the values reported by D0 \cite{D0} and CDF \cite{CDF}.
Small effects of absorption are by no means trivial in view of the large 
interaction cross section for many components of the
 hadronic wave function of the real 
photon. They may indicate that colour transparency
 is  at work here as the ZEUS  trigger may select point-like configurations
 in the photon wave function \cite{FSgap}.
  To check this idea it would be natural to
study the $A$-dependence of rapidity gap survival.
 It is demonstrated in \cite{FSgap} that this probability
 strongly depends on the effective cross section of the interaction
 of
the photon with the  
quark-gluon configurations involved in producing
 rapidity gap events. One would be sensitive to cross sections as small as
$ \sim$ 5 mb.

\end{enumerate}

\subsection{Parton Propagation in Nuclear Matter}
\label{partonprop}

\subsubsection{Introduction}
Measurements of final state hadrons  allow to investigate the 
effects of partons propagating through nuclear matter. In the present section
we discuss some of these possibilities. The discussion is restricted to {\it
incoherent} 
 phenomena (coherent nuclear interactions were 
discussed in previous sections).

There
are essentially two types of measurements which can be useful:
 (a) energy loss of high-energy particles and (b) increase of their transverse
momentum, both
studied as a function of the nuclear number A and/or number of nucleons emitted
 from the target nucleus. They are sensitive
to different aspects of the  interactions. Energy loss reflects the properties
of inelastic collisions: the value of the inelastic parton-nucleon
cross section and of the inelasticity. The increase of transverse momentum and
emission of the nucleons from the target
can be induced by  elastic as well as by inelastic collisions and thus can
provide information on both.

At this point we emphasize the importance of the measurement
of the distribution of the nucleons (protons and neutrons) emitted from the
target nucleus during or after the interaction. Such measurements give
direct access to the number of secondary interactions inside the target
\cite{aos}, as
was already realized (and used) in numerous emulsion experiments where 
protons with momenta 
$250  \le  p_N \le 700$ MeV/c  were measured
\cite{em}.
Related  information can be inferred from the  
measurement of the production of soft neutrons ($E_n \le 10 $ MeV) which was 
studied recently by the E665 collaboration \cite{e665}. Measurements
of the emitted nucleons and of their energy spectrum should thus be considered
a high priority. They allow to improve greatly the analysis of the data in 
three respects: (a) by studies of the distribution of nucleons themselves one 
obtains information of the strength of the secondary interactions and - more 
essentially- on the fluctuations which are expected to be large and would
otherwise hamper the interpretation of the data; (b) By studying the
interactions as a function of the number of the emitted nucleons
 one can obtain not only a larger lever arm in terms of the
number of secondary interactions but also, at a {\it fixed} nuclear number, a
clean sample, free of  possible biases related to the use of different
targets; (c) a really exciting possibility is to develop a trigger for events
with a large number of emitted nucleons (or highly reduced charge
of the nuclear remnant). This would allow to study in detail the rare
events with particularly strong secondary interactions.  Further 
quantitative studies (both theoretical and experimental) of this problem 
are necessary to clarify the relation between the emitted nucleons and 
number of secondary interactions.  Some work on these lines was already
presented during this workshop \cite{EGK,KRASNY,STZ}.  The 
detection of such nucleons at HERA will be simpler than in fixed target 
experiments since they will be boosted by the motion of the target nucleus
\cite{KRASNY}.  

As we discussed previously 
there should be a substantial difference between the nuclear
effects observed in the region of ``finite'' $x$ ($x$ greater
than, say, 0.05) and the region of very small $x$ ($x$ smaller than,
say, 0.001). The reason is the different nature of the photon-nucleon
interactions
in these two regions caused by the difference in life times
 of the relevant photon fluctuations into a $q\bar q$ pair.

One sees from eq.(~\ref{1}) that at finite
$x$ the life time of a fluctuation is rather short (smaller than 2 fm).
  This has two consequences: (a) the interaction of the photon must
take place inside one of the nucleons of the target (nuclear coherence
suppressed) and
(b) the high-energy part of the interacting photon
fluctuation  can be well approximated by a simple ``bare'' quark.
This can be seen as follows. The time necessary to produce a
high-energy gluon from a quark is given by
\begin{equation}
\tau_g = \frac{ 2E_g}{q_t^2}         \label{3}
\end{equation}
where $E_g$ is the gluon energy and $q_t$ its transverse momentum (with respect
to the quark). From the obvious condition $\tau_g < \tau$ we then obtain
\begin{equation}
E_g < \frac{\beta k_t^2}{2m_N x }.     \label{4}
\end{equation}
We conclude that at finite $x$
there is simply no time to produce the energetic gluon cloud.

For DIS
 nuclear collisions this picture implies that after the first interaction of
the virtual photon  the nucleus is penetrated by  one bare quark (following
approximately the direction of the virtual photon).
Alternatively one can consider the process of production of two high $p_t$
jets in {\it photon-gluon fusion}. In this case one studies propagation 
through the nucleus of a colour octet state.
  Thus the observed nuclear effects  measure interactions of a {\it bare}
parton 
 (or a system of {\it bare}
 partons)
 in  nuclear matter.

The situation is rather different in the region of small $x$. In this case
there is enough time for the $q\bar q$ system coupled to the virtual photon to
``dress'' itself into a cloud of energetic gluons and $q\bar q$ pairs
 (in the limit $x_{Bj} \rightarrow 0$ the condition (\ref{4}) is not
restrictive). Consequently,
the system traversing the nucleus is a complex
multiparton system resembling in some respects an ``ordinary'' hadron. The
observed nuclear effects measure interactions of this multiparton system ({\it
dressed} quark or $q\bar q$ dipole) in
 nuclear matter. It should thus  not be surprising that the expectations for
the region of small $x$ are rather different from those at finite
$x$.

\subsubsection{Current Experimental Situation}
Two major manifestations of the interaction of a parton propagating 
through the nuclear medium which were studied experimentally so far are the 
parton energy loss  and the broadening of its transverse momentum 
distribution.

The measurements of the leading hadron spectrum in deep inelastic 
 lepton-nucleus scattering have the biggest sensitivity to the energy loss.  
Measurements at incident energies below 50-100 GeV find a  
depletion of the leading particle spectrum which may be interpreted as due 
to the energy loss. At higher energies the effect nearly disappears  
\cite{E665h}, indicating that  energy losses for partons 
propagating through nuclear matter are definitely smaller than 
$\Delta E /dz \le 1$ GeV/fm.
This is in agreement with the observed low multiplicity of low energy 
neutrons in $\mu Pb$
interactions at high energies \cite{e665}, which is consistent with
knockout of one nucleon from lead \cite{STZ}. This makes it impossible 
to address directly the question of energy losses at high energies.

The phenomenological situation with  transverse momentum
broadening is somewhat confusing at the moment. The $\mu$-pair production 
experiments \cite{Alde} which measure the broadening of the 
{\it incident } quark find rather  small  broadening: $\Delta p^2_t \sim 0.1$ 
GeV$^2$
for the distance, $L \sim 5$ fm. At the same time the transverse momentum 
broadening for the jets produced by {\it outgoing}
partons in photon-nucleus interactions seems to be much 
larger \cite{gammaA}. Theoretically this difference is not understood 
\cite{LQS}.

 \subsubsection{Perturbative QCD Expectations for Finite $x$}

    All estimates of nuclear effects in this region of $x$ accept that the
interaction of a {\it bare} quark in nuclear matter is dominated by
colour-exchange processes which lead to break-up of the
``wounded'' nucleon in the target but do not slow down significantly the
energetic quark. This is based on the argument that the
quark in question can only emit  gluons satisfying the condition $\tau_g <
\Delta l$ where $\Delta l $ is the distance between the subsequent collisions.
 From eq.(\ref{3}) we deduce that the
 energy loss in one secondary collision is limited by
\begin{equation}
\Delta E < 2 q_t^2 \Delta l,          \label{5}
\end{equation}
 i.e., by a
value which is independent of the energy of the quark. Consequently, a high
energy quark can lose only a small fraction of its energy.
 A precise evaluation
of
the energy loss in a collision with one target nucleon is not possible  at the
 present stage of the theory.

However recently a significant progress 
was obtained in the analysis of the propagation of 
a virtual parton through the nuclear medium \cite{Baier}. It was 
demonstrated that the Landau-Pomeranchuk-Migdal effect in QCD  is
qualitatively different from the case of QED.  It was argued that for 
sufficiently large distances, $L$, traversed by a parton the process 
is dominated by perturbative QCD though the momentum transfers in the 
individual collisions are small.

A simple relation was found between the $p_t$ broadening and the energy loss
\begin{equation}
-{dE \over dL}={\alpha_s N_c\over 8}\Delta p^2_t(L),
\label{Ept}
\end{equation}
which corresponds to substantially smaller energy losses than those implied 
by the inequality in eq.(\ref{5}).

Probably the most striking prediction is the 
quadratic dependence of the energy loss on the traversed distance, $L$, 
as compared to the nonperturbative models where it is approximately
 proportional to $L$. Numerically the authors find for a quark
\begin{equation}
-\Delta E \simeq  2 GeV \left({L \over \mbox{ 10 fm}}\right)^2,
\end{equation}
neglecting the $x$ dependence of the nucleon gluon density
 (and a factor of $\sim 2$ larger energy loss for gluons). 
If this effect is included the $L$-dependence is even steeper.
The numerical coefficient is estimated here from 
 the information on the nucleon gluon densities and consistent  with 
eq.(~\ref{Ept}) if one uses experimental data on $p_t$ broadening of 
the $\mu$-pair spectrum \cite{Alde}.  
Alternatively, if one uses the $p_t$ information from $\gamma A$ data 
a much larger energy loss is predicted.  
However even in this case the expected  energy loss is too small to be observed
{\it directly} in DIS at collider energies.

The energy loss occurs (for realistic nuclei) via the emission of one or 
two gluons 
with energies $\sim \Delta E$. So it would be  the best to look for the 
energy loss effects by  studying the production of hadrons in the nucleus 
fragmentation region. Quadratic dependence  
on $L$ will be manifest in the $A$-dependence of the number of knocked 
out nucleons, as well as in the fluctuations of the number of emitted  soft
protons and neutrons in the case of heavy nuclei. Current HERA detectors 
have good acceptance for such nucleons.  

Broadening of the transverse momentum spectrum may be more easy  to
access. The transverse momentum of a
parton
increases as a result of multiple collisions. Since the  momentum transfers 
in the subsequent 
collisions are independent, the increase of the (average)  transverse
momentum squared is proportional to  $L$ and hence to the number of 
secondary collisions. One expects for the quark  
\begin{equation}
\Delta p_t^2 \simeq 0.2 \mbox {GeV}^2{L \over 10 {\mbox fm}},
\end{equation}
while for gluons broadening is about a factor of 2 larger.

A word of caution is necessary here. For 
 distances typical even for heavy nuclei
the average momentum transfer is rather small so application of perturbative 
QCD may be difficult to justify. Also it is not clear whether it is safe to 
interpolate from fixed target energies to collider energies assuming that 
the momentum transfer in individual collisions is energy independent.

To study these effects at HERA in a clean way one needs to consider 
processes dominated by relatively large $x \ge 0.1$. They include processes 
of dijet production similar to those studied at FNAL \cite{gammaA}. 
An advantage of HERA is that it would be
 possible to use information about the decay of the nucleus to check 
the correlation between transverse momentum broadening and the number of 
struck nucleons.

\subsubsection{Parton Propagation at Small  $x$}

This region is more relevant for HERA but, unfortunately, theoretical estimates
are difficult and uncertain
because  the system traversing the nucleus is fairly complicated. Hence 
its interaction with nuclear matter not easy to evaluate.
 Perturbative QCD leads to the simple prediction that at large $Q^2$ and large 
incident energies,  
due to QCD factorization, the spectrum of leading hadrons \cite{Mue82} is 
given by,  
\begin{equation}
{1 \over \sigma^{\gamma^*A}(x,Q^2)}{d \sigma^{\gamma^*+A \rightarrow h +X}(x,Q^2,z) \over dz}= f(z,\ln Q^2),
\end{equation}
which does not depend on $A$.
 (Energy losses we discussed above do not change the $z$ spectrum in this limit).
 However,
it is far from clear if perturbative methods are applicable at all - even
at large $Q^2$ \cite{Bj96}. 
The incoming system can experience
several soft interactions with the nucleons of the target nucleus. However, 
the AGK technique which can be used to relate nuclear shadowing with 
the  $A$ dependence of diffraction and the fluctuations of hadron 
production in the central region, does not allow any predictions for 
the $A$ dependence of the  spectrum of  
leading hadrons in the current fragentation region since 
these effects depend on the details of the virtual photon wave function.   
In view of these difficulties it is not possible to
give unbiased quantitative predictions and  we restrict ourselves
to a discussion of qualitative expectations and possible interpretations of
the future measurements.

As we have already mentioned, at small $x$ the virtual photon fluctuates
into a
$q\bar q$ pair a long time before it enters the nucleus.
 This has several
consequences.
 First, it opens the possibility of {\it coherent} phenomena in
which the nucleus participates as a whole which were discussed in a 
previous section. 
They should be carefully separated
from the incoherent interactions we are concerned with.
 Second, there is
enough time for this fluctuation to emit a large number of gluons and new
$q\bar q$ pairs {\it before} it enters the nucleus.
 The
cross-section of such a ``dressed'' fluctuation is generally fairly
large, leading to substantial ``shadowing'' effects, as  already discussed.  
Finally, the system produced in the first collision is by no
means a single bare quark but rather a multiparton conglomerate - result of a
quark-gluon cascade  of length equal to the available rapidity,
i.e. very long at small values of $x$. Studies
of interactions in nuclear matter provide an opportunity to obtain
information about this object.

The cascade origin of the system travelling through the nucleus implies
strong correlations between partons. Consequently, large fluctuations are
expected in its physical properties  and thus  also in the observable nuclear
effects. For example, the cross section for secondary interactions is expected
to vary widely from event to event. This variation may or may not be correlated
with the fluctuations of the parton multiplicity and energy
distributions. For a clean interpretation of the results it is
therefore crucial to obtain information on the number of secondary
interactions inside the nucleus which is the most important
parameter determining the strength of the interaction. Fortunately, as we
have already explained, this
information is accessible through measurements of the distribution of nucleons
(protons and neutrons) emitted by the colliding nucleus. They seem thus
crucial for the success of this investigation.

   The distribution of the number of collisions gives straightforward
information
on the fluctuations of the cross section for secondary interactions inside the
nucleus. Its change  as $Q^2$ increases from $0$ to the deep inelastic 
region should give information on the extent to which the structure of parton
bound states (i.e. hadrons) differs from that of the ``dressed'' $q\bar q$ pair.

\begin{figure}
\begin{center}
\vspace{-0.7cm}
\leavevmode
\epsfig{file=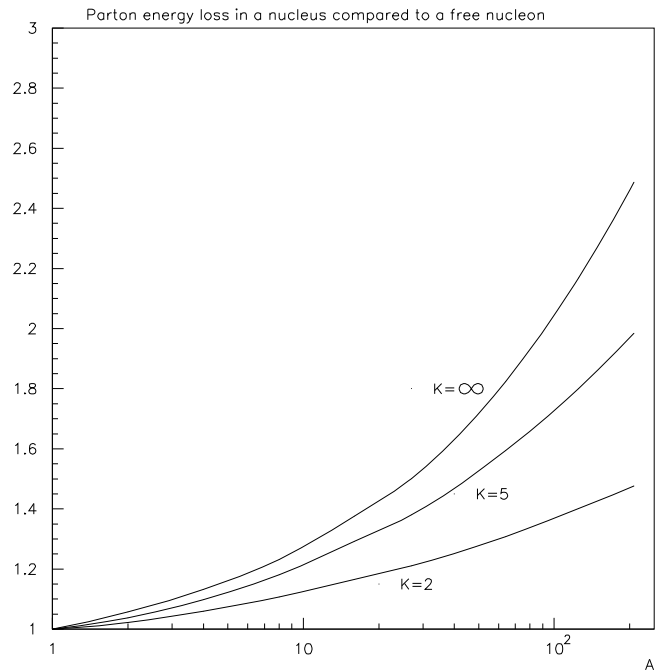,height=10cm}
\end{center}
\vspace{-1.2cm}
\caption{Energy loss of the incident virtual photon in nuclei assuming  
that the photon fluctuates into $K$ constituents interacting 
independently in nuclear matter with an interaction 
 cross section of 20 mb. }
\label{parton}
\end{figure}

Similar remarks apply to measurements of the energy loss (which measures 
the energy 
distribution inside the parton system in question). Correlation between the
observed energy loss and the number of collisions gives information on
clustering phenomena inside the parton system. It will be interesting to
look
for possible effects of constituent quarks at low $Q^2$ and to see when they
disappear as $Q^2$ increases. Fig.~\ref{parton} illustrates the 
possibility of such an investigation. It shows the ratio of the total energy
loss of the incident
virtual photon in the nucleus to that in collision with one nucleon, assuming
that the photon fluctuates into $K$  constituents which interact 
independently (losing energy) with the target. One sees that the curves
corresponding to different values of $K$ are substantially different. This
seems to give a chance to see possible effects of constituent quarks
($K=2$) at low 
$Q^2$ which should gradually disappear ($K \rightarrow \infty$) as $Q^2$
increases.  
It should be remembered, however, that variations in the cross-section of the
photon fluctuation in question will smear out this clean effect. It is
therefore necessary to rely on an independent estimate of the number of 
collisions as we already emphasized at the beginning of this section.  
In this case the ratio in question is given by 
\begin{equation}
\frac{\Delta E_A}{\Delta E_1} = K(1-(1-1/K)^N)   \label{6}
\end{equation}
 ($N$ is the number of collisions) and one can easily see that this is 
indeed a much better way to analyze the  data, particularly at high $N$.  

 It should be noted at this point
that, in contrast to the situation at finite $x$, in the present case the
energy loss is expected to be a finite fraction of total available energy, 
as explained above. Consequently, there seem to be no
particular difficulties in this measurement.

To summarize, one expects a dramatic increase of the energy
loss in nuclear matter when one goes from finite to  small values of $x$.
If confirmed by future data this should allow to study details of the
parton structure of the QCD cascade.  Investigation of the
transition region of $x \approx 0.01$ is also of great interest.  
HERA is very well suited for this task.

\subsection{ Connection to Heavy-ion Collisions at High Energies}

The interplay between the physics which can  be studied in high-energy
$eA$ collisions at HERA and that to be studied in the heavy ion physics
was discussed at the dedicated workshop ``Nuclei at HERA and Heavy Ion  
Physics'' which was held at Brookhaven National Laboratory in 
1995. It was concluded that the measurements 
 of $eA$ collisions at HERA can provide crucial information necessary for unambigous
interpretation of the heavy ion colllisions at RHIC and LHC for  establishing 
 whether a quark-gluon plasma is formed in these collisions. 

Three major links are

$\bullet$ {\it Nuclear gluon shadowing}

One needs $xg_A(x,Q^2)$ for $x\sim 10^{-2}$, $Q^2 \sim 1-10 $GeV$^2$ 
and  $x\sim 10^{-3}$, $Q^2 \sim 10 $GeV$^2$ to fix the initial conditions 
at RHIC and LHC respectively. This is especially important for the LHC 
since mini-jet production determines the initial conditions for 
$\sqrt{s} \ge 100 $GeV $\dot A$. The  bulk of the particles produced 
at central rapidities in $AA$  collisions at the LHC is expected to be 
generated due to this mechanism \cite{GW}.
Currently uncertainties in nuclear shadowing transform into at least 
a factor 2-4 differences in the final transverse energy flow \cite{G}.

$\bullet$ {\it Jet quenching}

Recent QCD studies \cite{Bqg} have demonstrated that the medium
induced  energy losses and $p_t$ broadening
of a high energy parton traversing a hot QCD medium 
are much larger than in the case of a cold medium.  This provides
a unique  new set of 
global probes of the properties of the state formed during $AA$ collisions
\cite{G}.
To interpret unambiguously this effect it is necessary both to measure the 
nuclear gluon shadowing and to study the parton propagation in cold matter  
in DIS to confirm that the energy losses ($p_t$-broadening) remain small at 
energies comparable to those to be studied at RHIC and LHC.  

 $\bullet$ {\it Testing of soft dynamics of interactions with nuclei}

Study of $eA$ interactions at HERA in {\bf the same energy range }
as that to be studied in $pA$ and $AA$ collisions at 
RHIC ($\sqrt{s} \sim 200$ GeV)
will provide a unique testing ground for the modern models of interactions 
with nuclei which aim at describing on the same footing 
$ep, eA, pp, pA, AA$ collisions \cite{EGK}.  It would allow to be established  
whether or not the same dynamics determines hadroproduction in 
$eA$ collisions and in central $AA$ collisions.

\section{Acknowledgements}

We thank all our colleagues in working group 8 for their invaluable 
assistance and for numerous discussions.   
AB acknowledges support by the KBN grant No 2 P03B 083 08 and by a PECO
grant from the EEC Programme ``Human Capital and Mobility'', Network  
``Physics at 
High Eneregy Colliders'', Contract No ERBICIPDCT940613.  MS thanks the U.S. 
Department of Energy for financial support under grant number 
DE-FG02-93ER-40771.

\end{document}